\documentclass[fleqn,usenatbib]{mnras}

\usepackage{newtxtext,newtxmath}

\usepackage[T1]{fontenc}
\usepackage{ae,aecompl}

\usepackage{graphicx}	
\usepackage{amsmath}	
\usepackage{amssymb}	
\usepackage{flexisym}	
\usepackage{gensymb}

\defcitealias{2016ApJ...831...89S}{S16}
\newcommand{\fgl}{J0427}

\title[Observations of 4FGL J0427.8-6704]{Optical, X-ray, and $\gamma$-ray observations of the candidate transitional millisecond pulsar 4FGL J0427.8-6704}

\author[M. R. Kennedy et al.]{M. R. Kennedy$^{1}$\thanks{Email: kennedy.mark@manchester.ac.uk}, R. P. Breton$^{1}$, C. J. Clark$^{1}$, V. S. Dhillon$^{2,3}$, M. Kerr$^{4}$,\newauthor
D. A. H. Buckley$^{5}$, S. B. Potter$^{5}$, D. Mata S\'{a}nchez$^{1}$, J. G. Stringer$^{1}$, T. R. Marsh$^{6}$
\\
$^{1}$Jodrell Bank Centre for Astrophysics, School of Physics and Astronomy, The University of Manchester, M13 9PL, UK\\
$^{2}$Department of Physics and Astronomy, University of Sheffield, Sheffield S3 7RH, UK\\
$^{3}$Instituto de Astrof\'isica de Canarias (IAC), E-38200, La Laguna, Tenerife, Spain\\
$^{4}$Space Science Division, Naval Research Laboratory, Washington, DC 20375-5352, USA\\
$^{5}$South African Astronomical Observatory, PO Box 9, Observatory 7935, Cape Town, South Africa\\
$^{6}$Department of Physics, University of Warwick, Coventry CV4 7AL, UK
}

\date{Accepted XXX. Received YYY; in original form ZZZ}

\pubyear{2019}

\begin{document}
\label{firstpage}
\pagerange{\pageref{firstpage}--\pageref{lastpage}}
\maketitle

\begin{abstract}
We present an optical, X-ray, and $\gamma$-ray study of 1SXPS J042749.2-670434, an eclipsing X-ray binary which has an associated $\gamma$-ray counterpart, 4FGL J0427.8-6704. This association has led to the source being classified as a transitional millisecond pulsar (tMSP) in an accreting state. We analyse 10.5 years of Fermi LAT data, and detect a $\gamma$-ray eclipse at the same phase as optical and X-ray eclipses at the >5$\sigma$ level, a significant improvement on the 2.8$\sigma$level of the previous detection. The confirmation of this eclipse solidifies the association between the X-ray source and the $\gamma$-ray source, strengthening the tMSP classification. However, analysis of several optical data sets and an X-ray observation do not reveal a change in the source's median brightness over long timescales or a bi-modality on short timescales. Instead, the light curve is dominated by flickering which has a correlation time of 2.6 min alongside a potential quasi-periodic oscillation at $\sim$21 min. The mass of the primary and secondary star are constrained to be $M_1=1.43^{+0.33}_{-0.19}$ M$_{\odot}$ and $M_2=0.3^{+0.17}_{-0.12}$ M$_{\odot}$ through modelling of the optical light curve. While this is still consistent with a white dwarf primary, we favour the transitional millisecond pulsar in a low accretion state classification due to the significance of the $\gamma$-ray eclipse detection.
\end{abstract}




\section{Introduction}
``Redbacks'' are binary star systems which have a neutron star (NS) primary and a low-mass, near-main sequence companion. The neutron stars in these systems are detectable at radio and $\gamma$-ray wavelengths as millisecond pulsars (MSPs). In recent years, 3 ``redback'' systems have become increasingly important in understanding the evolution of MSPs in binary systems: PSR J1023+0038 (\citealt{2009Sci...324.1411A}; \citealt{2013ATel.5513....1S}), IGR J18245-2452 \citep{2013Natur.501..517P} and PSR J1227-4853 \citep{2014MNRAS.441.1825B}. These three systems have been observed to transition between a radio loud state, where the pulsar is detectable at radio wavelengths and there is no evidence for active accretion from the secondary, and a low-mass X-ray binary (LMXB) state, where emission from the radio pulsar is quenched and material flows from the secondary through the inner Lagrange point towards the NS primary, where it builds an accretion disc. 

There are 4 further systems which have been proposed to belong to the same group as above based on their optical and X-ray behaviour: XMM J174457-2850.3 \citep{2014ApJ...792..109D}, 3FGL J1544.6-1125 \citep{2017ApJ...849...21B}, 3FGL J0427.9-6704 \citep[][hereafter referred to as S16]{2016ApJ...831...89S}, and CXOU J110926.4-650224 \citep{2019A&A...622A.211C}. These 7 systems make up the transitional millisecond pulsar (tMSP) class of interacting binaries.

This paper focuses on one of the candidate tMSP systems, 1SXPS J042749.2-670434, which is a binary system with an 8.8 hour orbital period. The X-ray source has been associated with the bright $\gamma$-ray source 3FGL J0427.9-6704. Since the publication of the original $\gamma$-ray association by \citetalias{2016ApJ...831...89S}, the \textit{Fermi} LAT 8-year Source Catalog has been released (4FGL; \citealt{2019arXiv190210045T}), and the $\gamma$-ray source has been renamed 4FGL J0427.8-6704. This association has been made due to the presence of deep X-ray and optical eclipses in the light curve of 1SXPS J042749.2-670434 source which potentially coincide with an eclipse of the $\gamma$-ray source \citepalias{2016ApJ...831...89S}, with the $\gamma$-ray eclipse only detected at the 2.8$\sigma$ level. Hereafter 1SXPS J042749.2-670434 and 4FGL J0427.8-6704 are assumed to be the same source, and are collectively referred to as \fgl. Due to the lack of an accurate mass measurement, questions remain over whether the primary star in \fgl\ is a white dwarf (making the system a cataclysmic variable) or a neutron star/black hole (making the system a low-mass X-ray binary).

Based on the presence of $\gamma$-ray emission from the binary, and by estimating the primary mass using optical spectroscopy and photometry of the secondary star, \citetalias{2016ApJ...831...89S} have suggested that the primary star is likely an MSP, and that the system is a tMSP in the accreting state. However, at the time of publication, there have been no dedicated radio observations of \fgl\ reported in the literature, while a positive radio detection of the source would strengthen the classification of the system as a tMSP. Mudding the waters further is the conclusion by \citetalias{2016ApJ...831...89S} that the mass constraints on the primary are not strong, and that a white dwarf primary could still be possible, potentially making this system a member of the cataclysmic variable (CV) family of interacting binaries, which have white dwarf primaries. If this were true, the main issue would then be explaining the $\gamma$-ray emission from the source, as no known CV has a GeV $\gamma$-ray counterpart. 

Here we present high time-resolution optical photometry of \fgl\ taken in $u_s$, $g_s$, and $i_s$ filters simultaneously, X-ray data taken using \textit{XMM-Newton}, optical data taken using \textit{TESS} over 11 months, and 10.5 years of \textit{Fermi} data in an attempt to strengthen the detection of a $\gamma$-ray eclipse. 

The optical photometry reveals rapid flickering associated with an accretion disc, and tantalising hints of a periodicity close to $\sim$20 min, which we investigate using a combination of traditional timing analysis and Gaussian process modelling. The underlying orbital modulation coupled with the recent distance estimate to the system measured by \textit{Gaia} suggests that the primary is a neutron star, but the classification of \fgl\ as a tMSP in an accreting state still hinges on the observed $\gamma$-ray emission from the source, with no other evidence to support the classification as a tMSP over a regular low-mass X-ray binary.

\section{Observations}

\subsection{ULTRACAM}

\fgl\ was observed on 2017-10-14, 2017-10-16, 2017-11-22, and 2019-02-27 using ULTRACAM \citep{2007MNRAS.378..825D} mounted on the 3.59 m New Technology Telescope (NTT) at the La Silla Observatory. Simultaneous Super-SDSS g (\textit{g$_{s}$}) and i (\textit{i$_{s}$}) data were obtained with a typical cadence of 10 s, while simultaneous Super-SDSS u (\textit{u$_{s}$}) data were obtained with a cadence of 40 s (2017-10-14), 20 s (2017-10-16), and 10 s (2017-11-22, 2019-02-27), with only 24 ms dead time between individual exposures. The Super-SDSS filters are a set of filters which cover the same wavelength ranges as traditional SDSS filters but with a higher throughput (\citealt{2016SPIE.9908E..0YD}; \citealt{2018SPIE10702E..0LD}). These observations comprised of times on target of 0.61 h and 0.78 h, 6.16 h, 3.02 h, and 1.57 h respectively, with two runs on the night of 2017-10-14. 

The data were reduced using the ULTRACAM pipeline as described in \citep{2007MNRAS.378..825D}. The magnitudes were calibrated using previously determined zero points in each filter. Our band calibration is good to $\sim 0.1$ mag in the $i_{s}$ and $g_{s}$ bands based on the measured magnitudes of two nearby comparison stars. Unfortunately there are no measured SDSS $u_{s}$ magnitudes for any star in the field, but based on the accuracy of the $g_{s}$ and $i_{s}$ calibration we estimate the maximum band calibration uncertainty in $u_{s}$ to be $\sim 0.1$ mag.

Throughout the remainder of the text, any mention of $u_s$, $g_s$, or $i_s$ refers to data taken with ULTRACAM.

\subsection{HIPPO}

Photopolarimetric observations of \fgl\ were obtained on 2018-10-03 (2.4h), 2018-10-04 (1.9h), 2018-10-05 (2.9h), 2018-10-06 (2.6h) and 2018-10-07 (2.5h) using HIPPO \citep{2010MNRAS.402.1161P} in its all-Stokes mode. HIPPO was mounted on the 1.9m telescope of the South African Astronomical Observatory. Data were recorded at the default 1ms cadence and binned (10s photometry and 300s polarimetry) for analysis. The 2018-10-03, 2018-10-04 and 2018-10-05 observations were unfiltered whereas a broad band OG570 filter was used for the 2018-10-06 and 2018-10-07 observations. Data reduction proceeded as outlined in  \citet{2010MNRAS.402.1161P}.

\subsection{\textit{TESS}}
\begin{figure*}
	\includegraphics[width=0.99\textwidth]{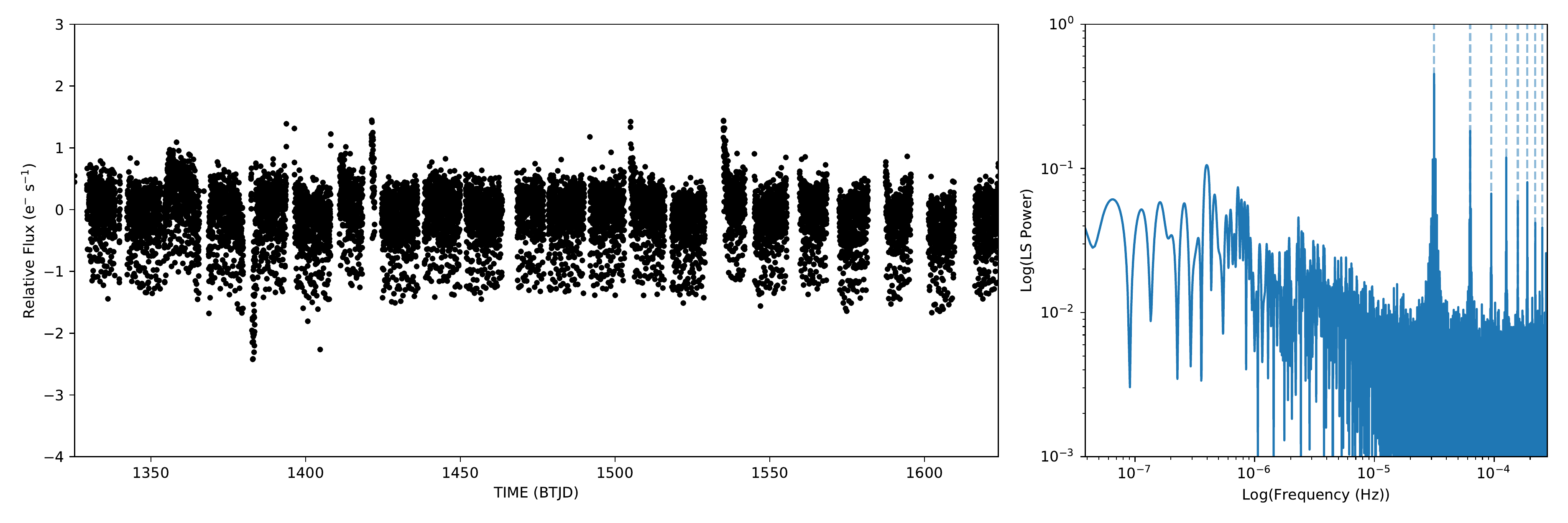}
    \caption{\textit{Left}:The extracted light curve for \fgl\ from \textit{TESS} Sectors 1-11. The aperture flux has been background subtracted, median subtracted, and $3.5$ $\sigma$ clipped. The amplitude of the variation is $\sim1.5$ e s$^{-1}$. Time is given in Barycentred \textit{TESS} Julian Date (BTJD), which is BJD-2457000.0. Gaps in the data are due to gaps in data acquisition between sectors.} \textit{Right:} A Lomb-Scargle periodogram of the \textit{TESS} data. The dashed lines mark the known orbital frequency and its first 8 harmonics.
    \label{fig:TESS_lc}
\end{figure*}

\fgl\ has been observed by CCD 4 of the \textit{Transiting Exoplanet Survey Satellite} (\textit{TESS}) during each Sector of Cycle 1 up until the submission of this paper (S001-S011). \textit{TESS} records full frame images every 30 mins during a Sector, with observations of each Sector lasting for 28 days. The data are taken through a wide filter which covers 6000-10000 \AA\ ($\lambda_{\rm{c}}=7865$\AA). The \textit{TESS} cut out images around \fgl\ were downloaded for all available Sectors. Extraction of a calibrated light curve was not possible, as the 21'' pixel size of \textit{TESS} means that \fgl\ is blended with several sources in the images. We constructed custom source and background apertures for each Sector, extracted the source aperture flux and removed background variations and flagged data where the background exceeded 100 electrons s$^{-1}$ pixel$^{-1}$, and then subtracted the mean flux value from the source light curve such that the residual light curve only showed variation in the source aperture. Even after this procedure, there were still many outlier points in the light curve. To remove these, we $\sigma$-clipped the data, setting $\sigma=3.5$, removing a further 606 of the 12545 data points. The resulting light curve over all 11 sectors is shown in Figure~\ref{fig:TESS_lc}, and the amplitude of the variation was $\sim1.5$ e s$^{-1}$. There are still clear systematics in this light curve, particularly at the start and end points of each Sector.

\subsection{\textit{XMM-Newton}}
\fgl\ was observed by the \textit{XMM-Newton} spacecraft starting 2017-05-02 16:14:17 and ending 2017-05-03 13:45:57 UTC for a total observation length of 77.5 ks, covering a total of 2.4 orbital periods. The European Photon Imaging Camera (EPIC) -pn \citep{epic_pn}, -MOS1, and -MOS2 \citep{epic_mos} CCDs were all operated in Full Frame mode with a medium filter inserted. The Optical Monitor (OM; \citealt{xmm_om}) was operated in fast mode, with a white filter inserted. While both Reflection Grating Spectrographs \citep{xmm_rgs} were operational, these data will not be discussed as no appreciable signal was detected.

\begin{figure}
	\includegraphics[width=0.99\columnwidth]{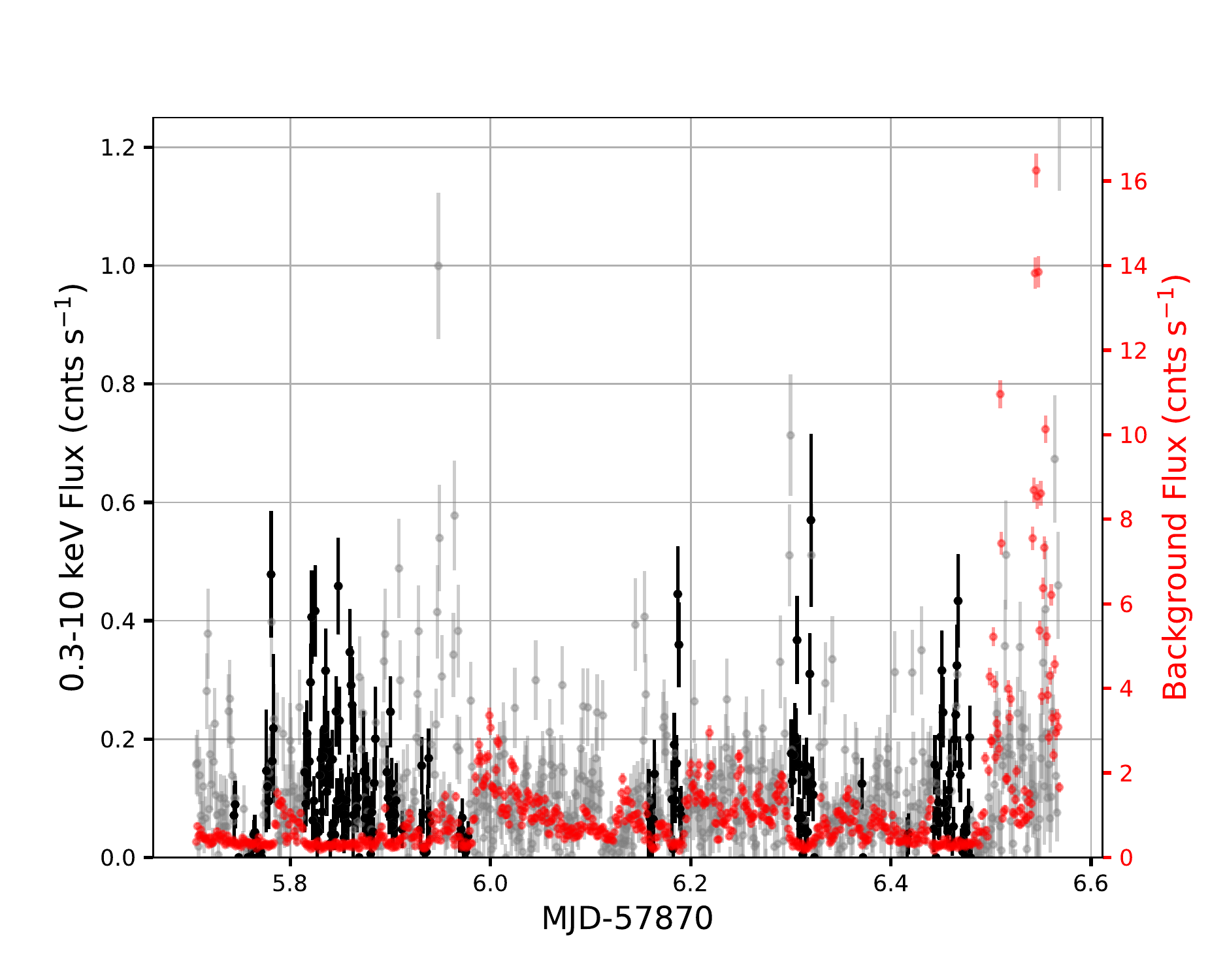}
    \caption{The extracted X-ray light curve of \fgl\ (black and grey) alongside the high energy background light curve (red). The grey X-ray data were excluded from spectral analysis due to the high background during these times.}
    \label{fig:xray_lc}
\end{figure}

The data were reduced using the Science Analysis Software (SAS) v16.1.0. The PN and MOS data were processed using the SAS commands \texttt{Epproc} and \texttt{Emproc} respectively. Unfortunately, the 77.5 ks exposure suffered from periods of severe high background. Figure~\ref{fig:xray_lc} shows the 0.3-10 keV light curve alongside the high energy background light curve which is used to characterise the intensity of soft proton flaring during observations with the EPIC instruments. Any period when this flaring activity is above 0.5 counts s$^{-1}$ ($\sim75$\% of the observation duration)  had to be discarded for spectral analysis. All of the X-ray data were considered when looking for X-ray eclipses in the X-ray light curve.

\subsection{\textit{Fermi} LAT}
For the analyses reported below, we selected Pass 8 \citep{2013arXiv1303.3514A} data from the \textit{Fermi} Large Area Telescope \citep{2009ApJ...697.1071A}, collected between 2008-08-04 and 2019-01-08 with reconstructed energies 0.1$<E<$30\,GeV, reconstructed positions lying within 2\degr of \fgl, and with an event class of 128 \footnote{In line with the chosen instrument response function of P8R3\_SOURCE\_V2, see \url{https://fermi.gsfc.nasa.gov/ssc/data/analysis/documentation/Cicerone/Cicerone_LAT_IRFs/IRF_overview.html}}. The associated source is modeled with a log parabolic shape, and we use the source list and associated diffuse models to compute the probability \citep[``weight'',][]{2011ApJ...732...38K} that each photon is associated with the FL8Y counterpart or with a background source.

\section{Phased light curve}	
\begin{figure*}
	\includegraphics[width=0.95\textwidth]{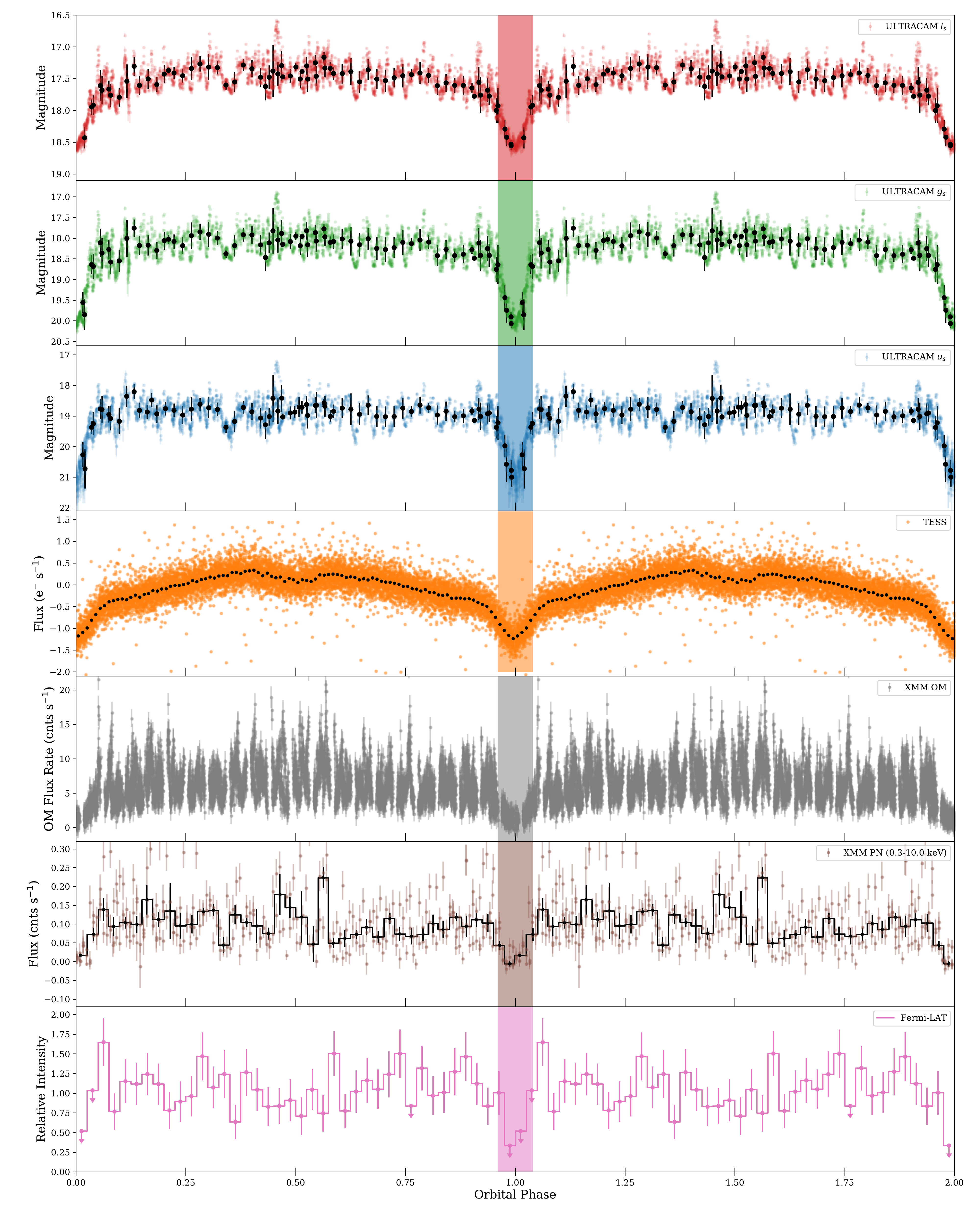}
    \caption{The phased light curve of \fgl\ in $i_s$, $g_s$, $u_s$ bands taken with ULTRACAM (top 3 plots), the \textit{TESS} data (orange, middle), the \textit{XMM-Newton} OM data (grey, 3rd from bottom), 0.3-10 keV X-ray data (2nd from bottom, brown), and \textit{Fermi} LAT data (pink, bottom, upper limits marked). Overlap from the 3 nights of ULTRACAM observations obscures some of the flaring activity around orbital phase 0.6. Orbital phase has been repeated for clarity. The black data points in the ULTRACAM data show the binned light curve which was used for modelling in Section~\ref{sec:orbit}, while the black points in the \textit{XMM-Newton} data are the raw data binned with a bin width of 0.025 in orbital phase. The shaded region highlights the eclipse data which were excluded when performing the timing analysis in Section~\ref{sec:ACF}.}
    \label{fig:phased_lc}
\end{figure*}

Figure~\ref{fig:phased_lc} shows the 3 bands of ULTRACAM data, the \textit{TESS} data, the X-ray light curve from the EPIC-pn instrument in the 0.3-10 keV band, the OM light curve, and the \textit{Fermi} LAT light curve (which is based on an analysis of \textit{Fermi} data which wll be discussed in Section~\ref{sec:gamma}), all phased to the orbital period using the ephemeris given in \citetalias{2016ApJ...831...89S} of 
\begin{equation}\label{eq:ephemeris}
T_{mid} (BJD) = 2455912.83987(95) + 0.3667200(7) \times E,
\end{equation}
where $T_{mid}$ is the predicted time of mid eclipse in Barycentric Julian Date and $E$ is the orbit number, with cycle 0 occurring at BJD 2455912.83987. The light curve shows a deep eclipse at phase 0 in each of the data sets, and lasts for 0.08 orbits ($\sim$42 min). Outside of eclipse, the optical and X-ray light curves are dominated by rapid flickering which occurs on a timescale of minutes. This variability is harder to detect in the \textit{TESS} data, as these data had a cadence of 30 min. Aside from the eclipse and flickering, the ULTRACAM and \textit{TESS} data show a hint of curvature on the orbital timescale which is strongest in the \textit{TESS} and $i_s$ data and is undetectable in the $u_s$ data. The \textit{TESS} data show evidence of a secondary eclipse at phase 0.5, which becomes more obvious after the data have been binned with a bin width of 0.01 in orbital phase. The X-ray eclipse has a similar duration to that of the optical eclipse, and the X-ray light curve is consistent with 0 flux being observed during the eclipse, suggesting the entire X-ray emitting region is being eclipsed.

\subsection{$\gamma$-ray Eclipse}\label{sec:gamma}
\citetalias{2016ApJ...831...89S} reported tentative evidence for a $\gamma$-ray eclipse at the same orbital phase as the eclipses observed in the optical and X-ray bands. To verify this claim, we folded the \textit{Fermi} photon timestamps with the \fgl\ orbital period and set the zero phase to fall on BJD 2457527.69139; this epoch is advanced by 0.25 in phase relative to that of \citetalias{2016ApJ...831...89S}, so that the eclipses fall at $\phi=0$.  We modelled the eclipse (or excess) using a top hat model such that within an eclipse of width $\theta$, the relative source rate is $\alpha$.  To enforce an average intensity of unity, the rate outside of the eclipse is thus $(1-\alpha\theta)/(1-\theta)$.  With this eclipse model, the Poisson likelihood is
\begin{align}
    \log\mathcal{L} = & \sum_{i\in\Theta}\log \left(w_i\alpha + 1-w_i\right) + \sum_{i\in\bar{\Theta}}\log \left(w_i \frac{1-\alpha\theta}{1-\theta} + 1-w_i\right) \\ \nonumber &-S\left(\alpha\eta_{\Theta} + \frac{1-\alpha\theta}{1-\theta}\eta_{\bar{\Theta}}\right),
\end{align}
where $\Theta$ indicates the phases of eclipse and $\bar{\Theta}$ the complement, $S$ is the total expected source counts ($S\approx\sum_iw_i$), and $\eta_{\Theta}$ represents the fraction of the instrument exposure falling in the eclipse.  We can approximate the eclipse shape as a truncated Fourier series, and \cite{2019arXiv191000140K} discusses the efficient evaluation of such likelihoods using Fast Fourier Transforms, allowing us to evaluate the likelihood over a wide range of trial orbital frequencies.  We perform this search using a 40-term Fourier series to approximate the eclipse profile and search over eclipse width ($0.01<\theta<0.5$), position ($0\leq\theta_0<1$), and amplitude ($\alpha>0$), taking the maximum value of the likelihood for each trial frequency. Compared to a uniform signal, we find $\delta\log\mathcal{L}=13.4$ for an eclipse of amplitude $\alpha=0.01$ and width $\theta=0.064$ centred at $\theta_0=1.000$.  Using this shape, but allowing $\alpha$ to vary, we searched $10^5$ neighboring orbital frequencies and found no signals with $\delta\log\mathcal{L}>8.8$, indicating a lower limit on the chance probability of $\sim$$10^{-5}$.  Moreover, because there is a single degree of freedom ($\alpha$), $\delta\log\mathcal{L}$ should follow a $\chi^2_1$ distribution, in which case the chance probability of observing $\delta\log\mathcal{L}=13.4$ is $2.2\times10^{-7}$, indicating the eclipse has $>5\sigma$ significance.  Moreover, the shape and phase are consistent with eclipses observed at lower frequency.

\begin{figure*}
	\includegraphics[width=0.99\textwidth]{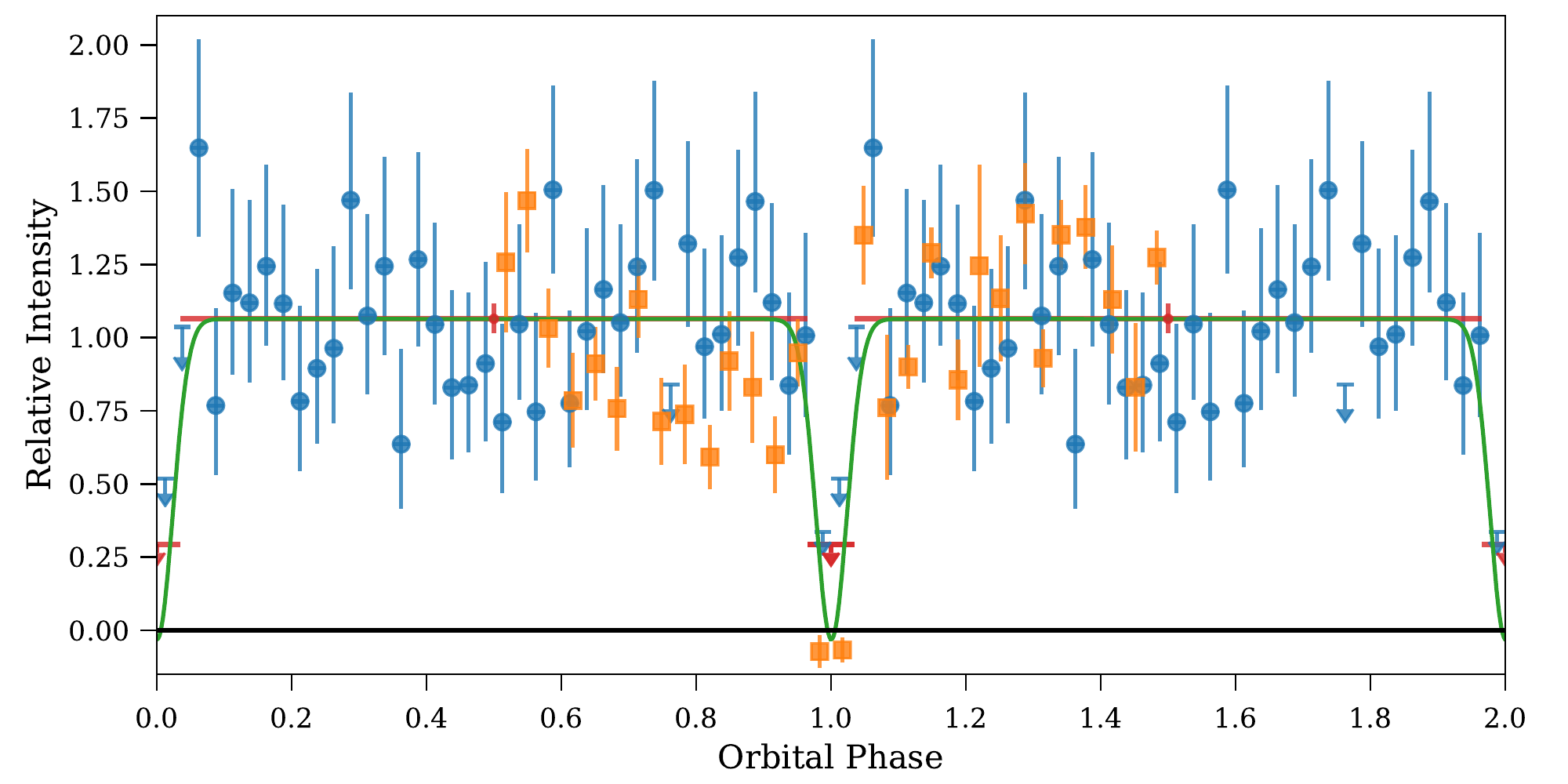}
    \caption{The $\gamma$-ray light curve obtained using \textit{Fermi}-LAT data and following the methods described in Kerr (2019, submitted).  The 40 blue points give maximum likelihood estimators for the intensity and its uncertainty, or upper limits, while the red markers give the same estimates for the two intervals determined with the Bayesian Blocks algorithm. For these two intervals, the extent on the x-axis indicates the interval boundaries, while the y-axis values give the intensity and uncertainty. The green line represents the maximum likelihood result from fitting a Gaussian eclipse model directly to the photon phases and weights.  Finally, for comparison, the orange points show NuSTAR data, folded on the \fgl\ orbital period as described in \citet{2016ApJ...831...89S}, scaled to give similar intensity to the $\gamma$-ray data.}
    \label{fig:fermi_lc}
\end{figure*}

To further characterize the eclipse, we first performed a simple maximum likelihood analysis to determine the relative source flux at 40 orbital phase bins, following the methods of Kerr (2019, submitted), in which the photon weights are used to approximate a full time-domain likelihood analysis.  The measurements (with $1\sigma$ errors and 95\% confidence upper limits) are shown in blue in Figure \ref{fig:fermi_lc}, clearly show the eclipse.  To identify any sharp features, we also apply a Bayesian blocks algorithm \citep{2013ApJ...764..167S} using 1000 orbital phase bins and the Poisson likelihood.  For a wide range of priors on the number of change points, the algorithm finds only two significant intervals, shown as the red levels in Figure \ref{fig:fermi_lc}, indicating a strong preference for a deep eclipse with sharp edges.  To confirm this, we directly fit the unbinned likelihood to find the best-fit values for ingress, egress, and flux within the eclipse, obtaining $0.963-1.041$ and $\alpha=0.13$.  The edges are relatively sharply constrained, with steep decreases in $\mathcal{L}$ indicating the left edge lies at $\phi\geq0.962$ and the right edge at $\phi\leq1.047$, while the flux during eclipse is only poorly determined, $0\leq\alpha\leq0.3$. The eclipse is apparently highly symmetric about the compact object superior conjunction, and the egress may be more gradual than the ingress.

To consider a more gradual eclipse, we modelled the eclipse as a Gaussian centred on $\theta_0$, $f(\phi)=1+A\exp\,[-(\phi-\phi_0)^2/2\sigma^2]$, and found maximum likelihood parameter values of $\theta_0$=1.001(6), $\sigma=0.023(5)$, and $A=-1.09(15)$.  The resulting model is shown as the green trace in Figure \ref{fig:fermi_lc}.  However, the improvement in $\delta\log\mathcal{L}=12.42$, compared to $\delta\log\mathcal{L}=15.10$ for the top hat model, suggests that the more rapid eclipse is the preferred model. Note that the likelihood calculation here uses an exact top hat representation, rather than the truncated Fourier series used in the frequency search reported above, yielding a slightly higher $\delta\log\mathcal{L}$.

\subsection{Orbital Period}
In order to try to improve the orbital period given by \citetalias{2016ApJ...831...89S}, we scaled the $i_s$ ULTRACAM data to match the amplitude of the variability in the \textit{TESS} data, and then computed a Lomb-Scargle periodogram (\citealt{1976Ap&SS..39..447L}; \citealt{1982ApJ...263..835S}) of the combined data set. The strongest peak was located at a period of 0.3667 d. We estimated an error on this period by minimising the $\chi^2$ value of a fit to the data using a SuperSmoother algorithm \citep{friedman1984smart}, as done in \citetalias{2016ApJ...831...89S}. The resulting best fit period was 0.36672$\pm$0.00002 d, which is not an improvement on the previous reported period. We also attempted to use Gaussian Process Modelling (which will be further discussed in Section~\ref{sec:GPM}) with a periodic kernel to determine an accurate period from the \textit{TESS} data. Such methods have proved effective in the past when dealing with data that contain flickering(\citealt{2017MNRAS.466.4250L}; \citealt{2018MNRAS.474.2094A}). However, when performed solely on the \textit{TESS} data, this method constrained the orbital period to be 0.366719(8) days, which is consistent with the previous reported error.

\subsection{Orbital modulation} \label{sec:orbit}
To better constrain the binary system parameters in \fgl, we modelled the $i_s$, $g_s$, and $u_s$ ULTRACAM data using the Eclipsing Light Curve (ELC) code \citep{2000A&A...364..265O}. Due to the rapid flickering in the light curve, we median binned each night of data with a bin width of 10 mins, with the error on each binned point taken to be the standard deviation of the points which make up that bin. This binned light curve is shown as the black points in the top three panels of Figure~\ref{fig:phased_lc}. The \textit{TESS} light curve was excluded from the modelling as the magnitudes could not be calibrated due to several stars lying within the aperture used for source extraction, while the OM light curve from \textit{XMM-Newton} was excluded as the data were taken using the white filter, which has a very large bandwidth.

There are many tunable parameters within ELC which are used to described the primary star, accretion disc, and secondary star. The variable parameters were the mass ratio $q=M_2/M_1$ (where $M_1$ is the primary mass and $M_2$ is the secondary mass), the system inclination $i$, the outer radius of the accretion disc $R_{\rm{out}}$, the temperature of the disc at the inner radius $T_{\rm{d}}$, the opening angle of the accretion disc $\beta$, the effective temperature of the secondary star $T_2$, the irradiation luminosity of the primary which is responsible for heating the secondary star $L_{\rm{X}}$, and the apparent radial velocity of the secondary star $K_2$. Note that $T_2$ is the average temperature over the entire surface of the secondary including the irradiated region of the secondary, meaning if irradiation of the secondary is high, then $T_2$ is likely to be much higher than the night side temperature of the star. 

In the following modelling, we assumed that the power-law exponent which controls the temperature profile of the disc (that is, $T(r) \propto T_d\left(\frac{r}{r_{\rm{in}}}\right)^{\xi}$ where $r_{\rm{in}}$ is the inner radius of the accretion disc) was $\xi=-0.425$, larger than what is expected from the typical $\xi=-0.75$ in ``steady-state'' accretion discs, but in-line with an irradiated accretion disc \citep{1981PASJ...33..365H} which is expected in this system due to the strength of the detected X-ray source. To ensure this choice of $\xi$ was not biasing our results, we also ran a Markov chain Monte Carlo (MCMC) as implemented as part of the ELC package (see \citealt{2004PhRvD..69j3501T}) with $\xi$ free. The only change this made to our final results were slightly higher errors on $T_{d}$, $T_{2}$, and $R_{out}$. During this modelling, $\xi$ quickly converged to $-0.4\pm0.1$, justifying the above assumption. Table~\ref{tab:model_lc_results} lists all of the relevant physical parameters used by ELC to model \fgl, and whether or not the parameter was fixed or allowed to vary.

All 3 optical bands of data were fit simultaneously. The best fit model was found by using a MCMC. We allowed 70 chains to evolve over 8000 steps each, discarded the first 250 steps of each chain as burn-in, and also filtered out models which had a $\chi^2$ which was more than 100 larger than our best fit model. The priors on all parameters were flat with hard edges at minima and maxima values, as given in Table~\ref{tab:model_lc_results}.

We also assumed a Gaussian prior on the radial velocity of the secondary star of $K_{2}=293\pm4$ km s$^{-1}$, in line with the measured value in \citetalias{2016ApJ...831...89S}. This $K_2$ value may be lower than the actual $K_2$ of the secondary in the system due to heating effects, but is still useful in providing a lower limit on the mass of the companion star.

\begin{figure*}
	\includegraphics[width=0.99\textwidth]{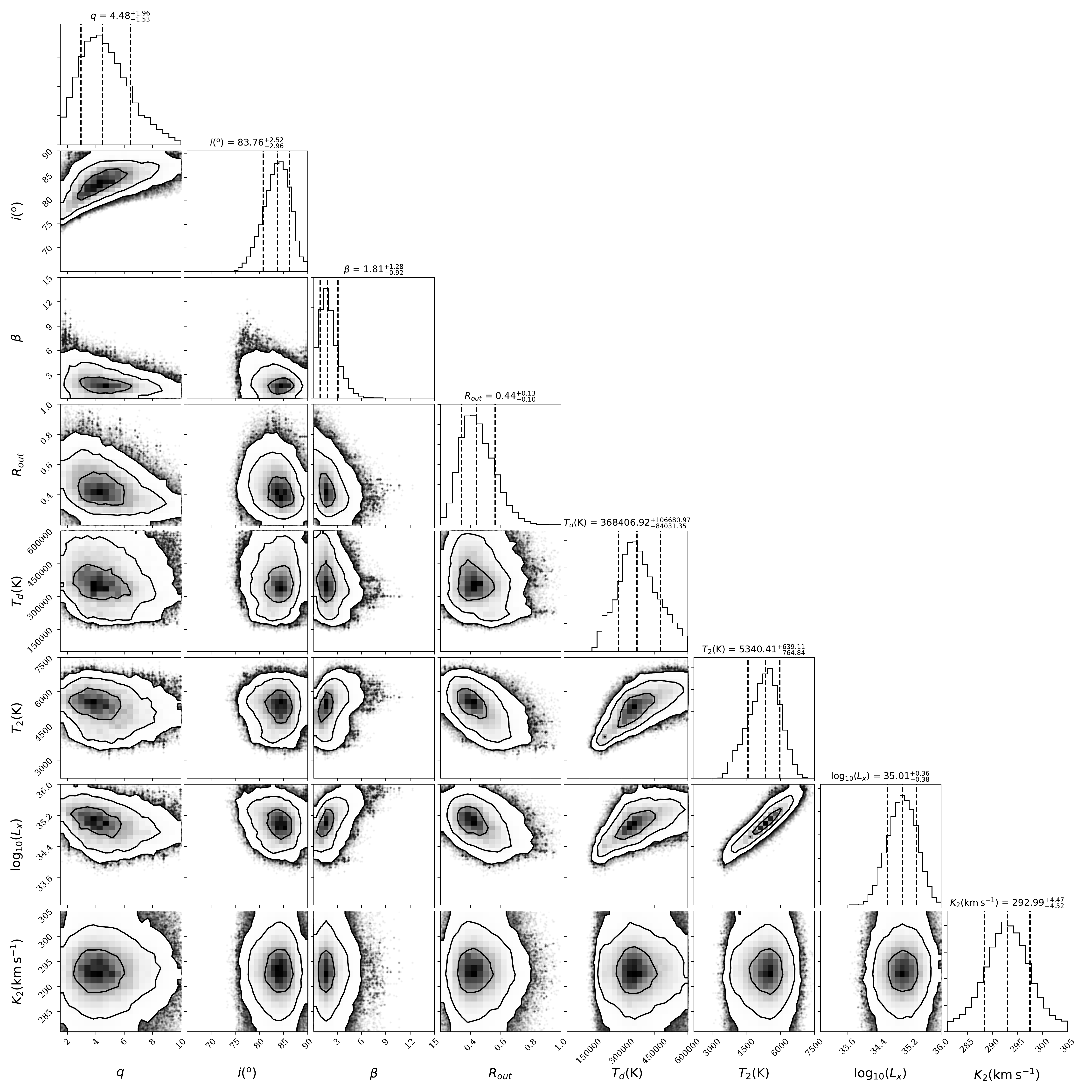}
    \caption{Corner plot from the MCMC analysis of \fgl\ using ELC. The priors for each parameter are given in Table~\ref{tab:model_lc_results}. There are clear correlations between $q$ and $i$, $q$ and $R_{\rm{out}}$, $T_d$ and $T_2$, $T_{\rm{d}}$ and $L_{\rm{x}}$, and $T_2$ and $L_{\rm{x}}$. We could not constrain $K_2$ better than what was done in \citetalias{2016ApJ...831...89S}. Dashed vertical lines mark the median value and 3$\sigma$ confidence intervals.}
    \label{fig:ELC_corner}
\end{figure*}

\begin{figure}
	\includegraphics[width=0.99\columnwidth]{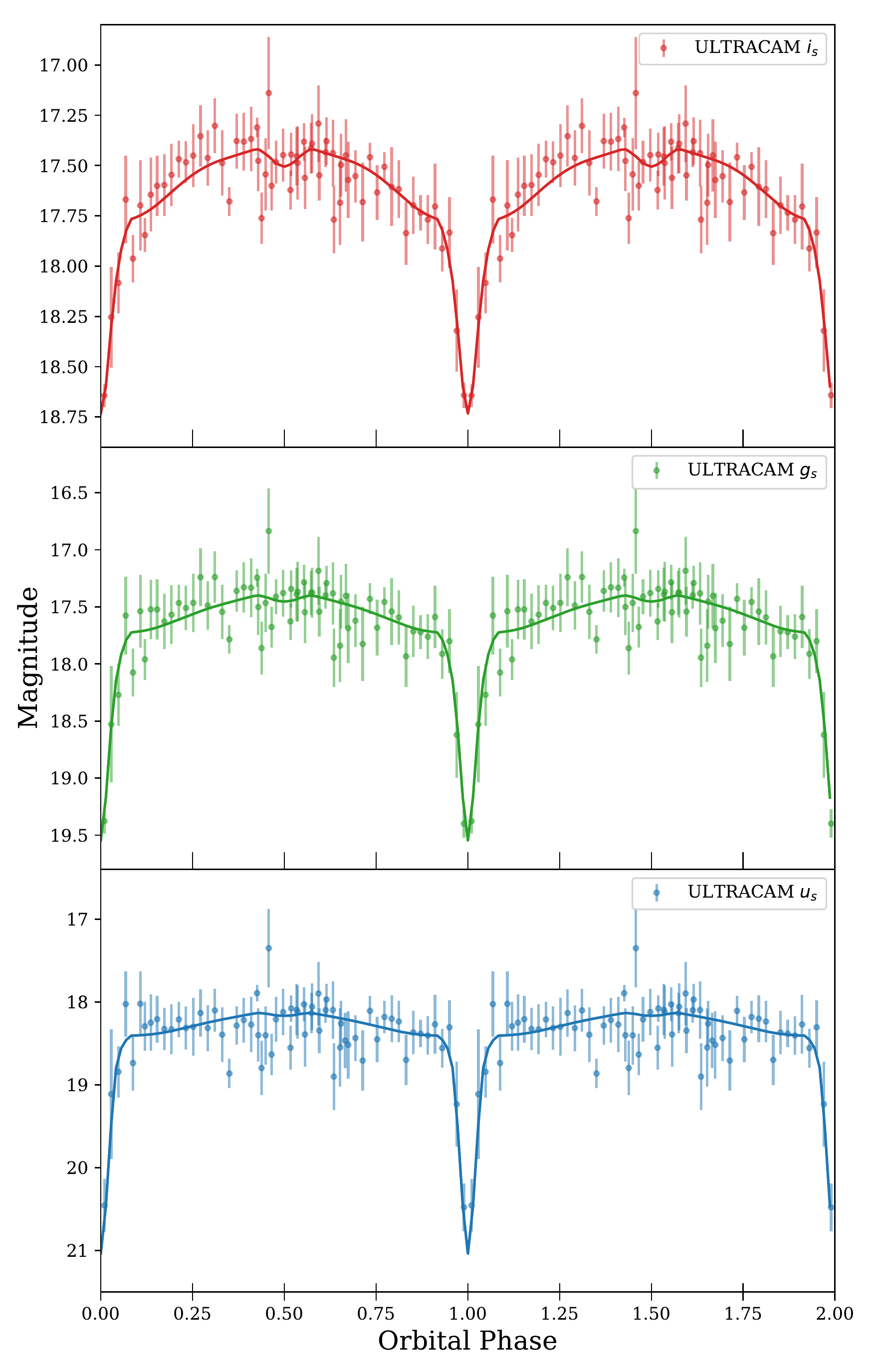}
    \label{fig:ELC_bestfit}
    \caption{Best fitting model generated using ELC (solid lines) alongside the 10-min median-binned data in each of the observed optical bands.}
\end{figure}

The final number of samples used for the following analysis was 680,642. The corner plot of the MCMC analysis is shown in Figure~\ref{fig:ELC_corner}, and our best fit model is shown alongside each optical band of data in Figure~\ref{fig:ELC_bestfit}. 
The corner plot shows that there are several degeneracies between various parameters ($q$ and $i$, $T_2$ and $L_{\rm X}$, $T_2$ and $T_{\rm{d}}$) and that several of the parameters are not well constrained using our data. Our final fit had a $\chi^{2}$ of 165.18 for 185 degrees of freedom, suggesting the error bars on the binned data points were over-estimated. The best constrained parameters are $i=84\pm3\degree$, $T_2=5300\pm600$ K, and $L_{\rm X}=\left(1\substack{+0.9 \\ -0.5}\right)\times10^{35}$ erg s$^{-1}$, where the errors have been scaled such that the $\chi^{2}_{\rm{R}}$ of the best fit model was 1. 

\begin{table}
	\centering
	\caption{Best fit values from the MCMC analysis of ELC. In the case where a parameter is given, the prior was assumed to be flat with the exception of $K_2$, which had a Gaussian prior with a $\sigma$ equal to the measurement error from \citetalias{2016ApJ...831...89S}. If a prior is not given, that means the parameter was fixed to this value. }
	\begin{tabular}{r c c c}
		\hline
							& Value						& Prior									& Ref\\
		\hline\hline
        $q$					& $5\pm$2					& 1.5-10.0								& \citetalias{2016ApJ...831...89S}\\
        $i$					& 84$\pm$3\degree			& 65\degree-90\degree\\
        $P_{\rm{orb}}$		& 0.3667200 d				& Fixed									& \citetalias{2016ApJ...831...89S}\\
        $R_{\rm{out}}^{\rm{a}}$ 	    & 0.4$\pm$0.1				& 0.2-1.0\\
        $T_{\rm{d}}$		& $(3.7\pm1.0)\times10^5$ K	& 1$\times10^5$-6$\times10^5$ K\\
        $\beta$				& 2\degree$^{+1}_{-0.8}$	& 0.01\degree-8\degree\\
        $T_{2}$				& $5300\pm700$ K			& 2800-6000 K\\
        $\log_{10}L_{\rm{X}}$	& $35.0\pm0.3^{\rm{b}}$				& 32.9$^{\rm{b}}$-36$^{\rm{b}}$								& \citetalias{2016ApJ...831...89S}\\
        $K_2$				& $293\pm5$ km s$^{-1}$		& $283-305$ km s$^{-1}$				& \citetalias{2016ApJ...831...89S}\\
		\hline
        \multicolumn{4}{l}{ $^a$ $R_{\rm{out}}$ is expressed as a fraction of the primary star's}\\
        \multicolumn{4}{l}{ effective Roche lobe radius, and so must be $<1$.}\\
        \multicolumn{4}{l}{ $^b$ units of $\log_{10}({\rm{erg}}\:{\rm{s}^{-1}})$}\\
	\end{tabular}
	\label{tab:model_lc_results}
\end{table}

\subsection{Colour information}
\begin{figure}
	\includegraphics[width=0.99\columnwidth]{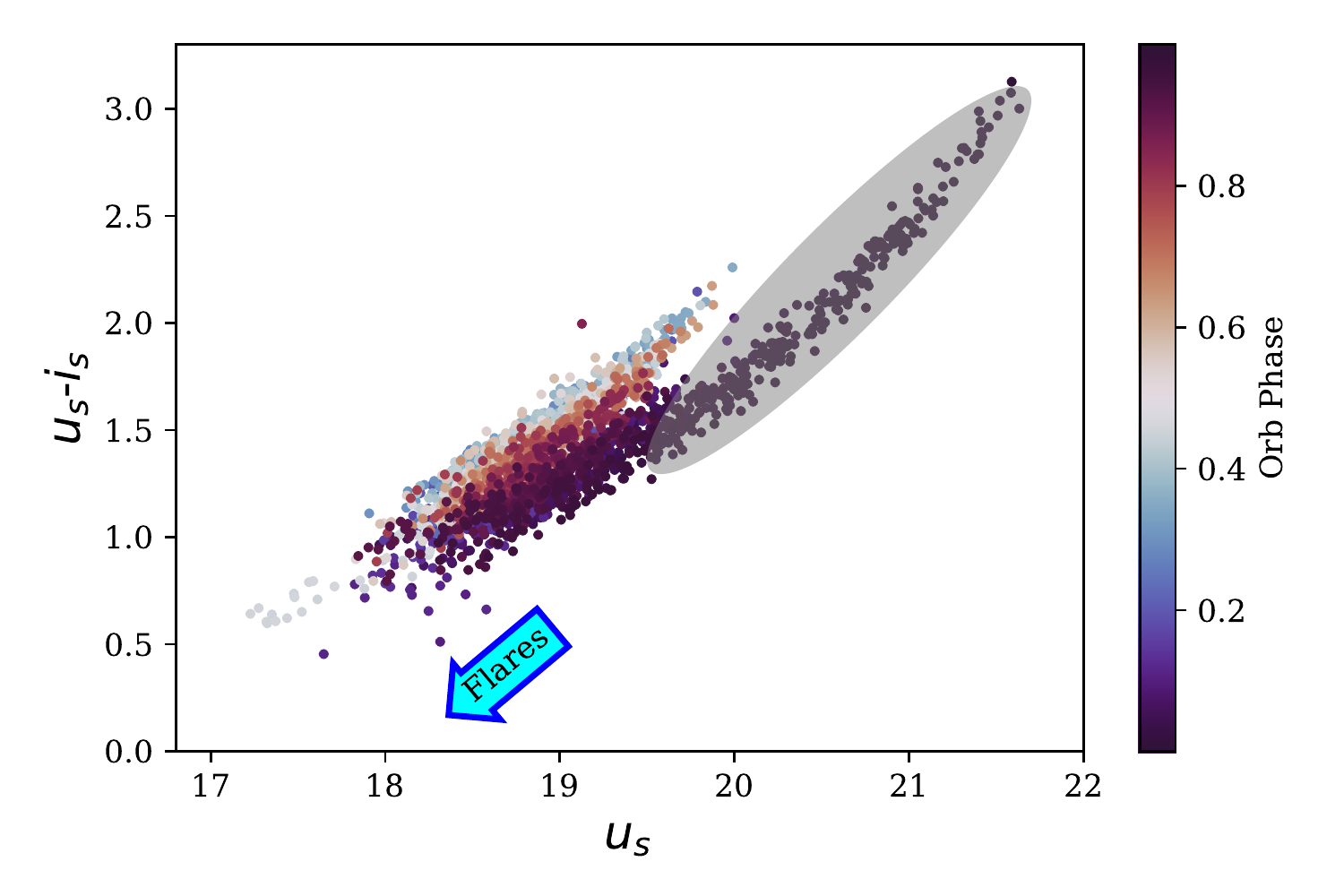}
    \caption{Colour-Magnitude diagram for the data of \fgl. The grey ellipse highlights data which were taken during eclipse, while the arrow points in the direction of the change in colour of the system due to flares. The colour scale indicates the orbital phase at which a given data point was taken, with deep red/blue corresponding to orbital phase 0.0/1.0 and lighter shades representing data taken close to orbital phase 0.5. The gradual reddening of the colour from phase 0.0-0.5 is from the curvature in the $i_s$ light curve.}
    \label{fig:colour_lc}
\end{figure}

Figure~\ref{fig:colour_lc} shows the $u_s$-$i_s$ colour versus $u_s$ magnitude for all of the simultaneous $u_s$ and $i_s$ ULTRACAM observations of \fgl. The points have been colour coded according to their orbital phase. There are 3 distinct features in the colour-magnitude diagram:

\begin{enumerate}
  \item A very rapid reddening of the object as the colour moves from the centre to the top right of the plot which occurs during the eclipse.
  \item A rapid, large variation in the colour which sees \fgl\ moving from the centre of the plot to the bottom left (the sources of these variations will be discussed in detail in a later section),
  \item A slow, small variation in colour due to the orbital motion of the secondary which sees the colour moving from the centre of the plot towards the upper left between orbital phases 0 and 0.5, and back down towards the bottom right from orbital phase 0.5 to 1.0.
\end{enumerate}

The gradual reddening of the object from orbital phase 0.0-0.5 and the reverse from phase 0.5-1.0 confirms the proposition made at the beginning of this section that the $i_s$ light curve has significant curvature over the orbital period while the $u_s$ light curve is relatively flat, and is line with the modelling discussed in the previous subsection.

\subsection{Flux distribution}
The ELC model found in Section~\ref{sec:orbit} was then subtracted from the $i_s$, $g_s$, and $u_s$ ULTRACAM light curves, such that the remaining signal only included noise and short time-scale flickering. We then generated histograms of the residual flux in each band to look for the bimodal distributions which have been seen in several tMSPs during their active states (\citealt{2015MNRAS.453.3461S}; \citealt{2017ApJ...849...21B}). The flux distributions only had a single peak with a high flux tail in each band. The same is true of both the \textit{TESS} and \textit{XMM-Newton} OM light curves.

There is a segment of data taken on 2017-10-16 during which the variability of the flux from \fgl\ decreased for $\sim 15$ min. The \textit{TESS} and \textit{XMM-Newton} OM data prove that this feature is not an orbital feature, as neither light curves show a drop in the average flux at the same orbital phase that this period of diminished variability was seen in the ULTRACAM light curve. This means that this drop was transient, and does not repeat every orbital period.

\begin{figure}
    \centering
    \includegraphics[width=0.99\columnwidth]{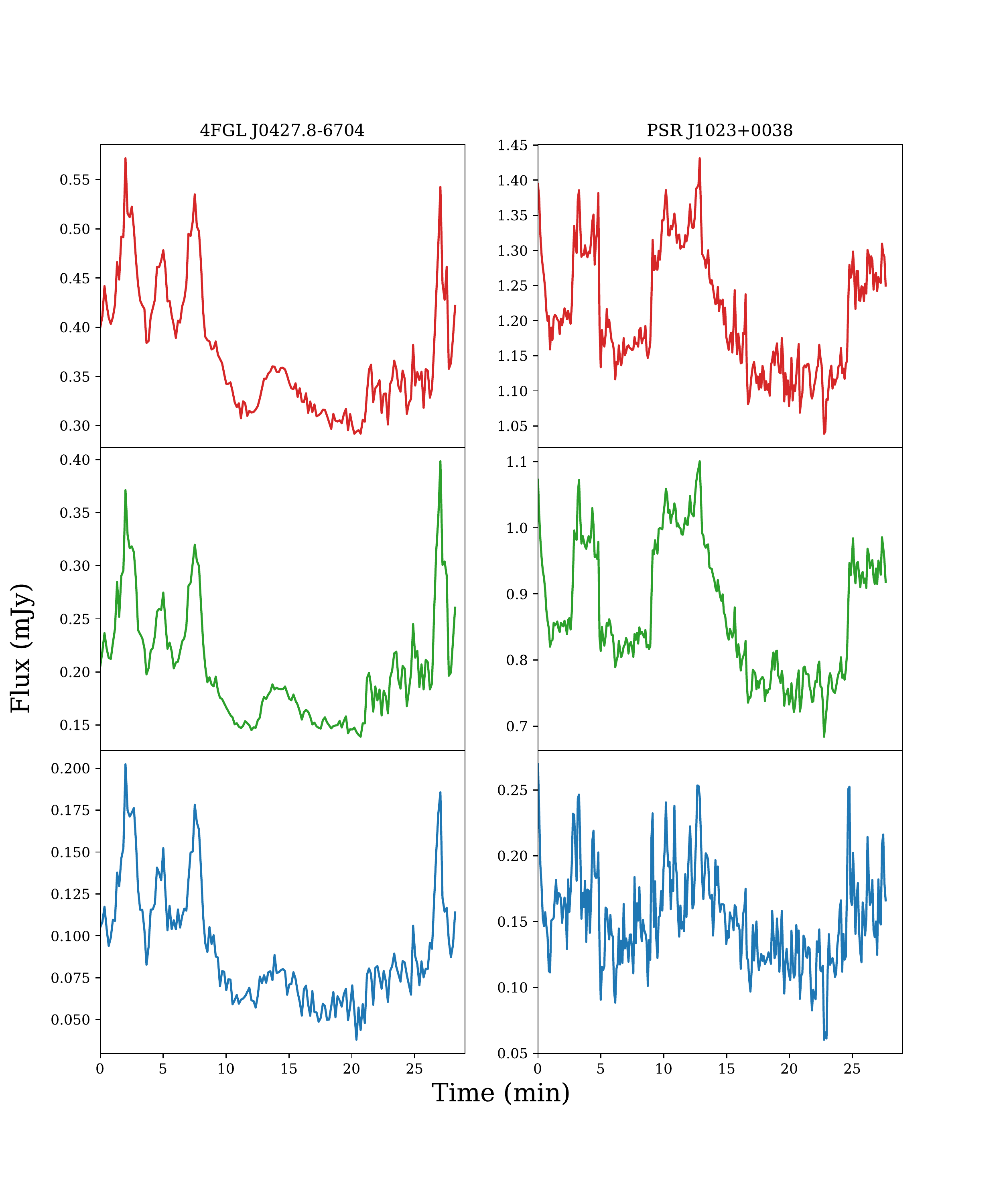}
    \caption{A comparison of the variability detected in \fgl\ (left) with PSR J1023$+$0038 (right), observed with ULTRACAM mounted on the NTT. Both data sets show similar behaviour, but the analogy is not perfect, with the high mode in PSR J1023$+$0038 showing plateaus that are not obvious in \fgl.}
    \label{fig:comp_lcs}
\end{figure}

Figure~\ref{fig:comp_lcs} shows this segment of data alongside a 29 min segment of the light curve of the tMSP PSR J1023$+$0038 which was also obtained with ULTRACAM when PSR J1023$+$0038 was in an accreting state on the night of 2019-03-01 using 10 s exposures. Both light curves show the same amplitude of variability. However, there are still differences in the light curves. In particular, the mode switching behaviour of PSR J1023$+$0038 has a very distinctive step shape, with clear plateaus during the high mode, while the features in \fgl\ more closely resemble flares rather than the bi-model behaviour seen in PSR J1023+0038.

\subsection{Photopolarimetry}

The photopolarimetric observations were binned to 300 s exposure times for analysis. We did not find any indication of linear polarization above the sky background. However, due to the faintness of \fgl\ and the moonlit linearly polarized sky, we were only able to place an upper limit of $\sim$ 3\% on the linear polarization of \fgl. Circular polarization displayed random excursions around zero percent, consistent with a non-detection with a limit of $<1$\%. 

\section{Short time-scale Variability}
While the orbital modulation dominates the long-term optical variability of \fgl\, there are also strong short-term flares visible in the data. We begin our discussion on this short-term variability by first considering the \textit{TESS} data (due to its long cadence), followed by the \textit{XMM-Newton} OM data, and concluding with the behaviour seen in the highest quality data, the ULTRACAM light curve. The 3$\sigma$ levels as plotted in the various periodograms in this section were derived as described in Appendix~\ref{sec:app_1}.

\subsection{TESS}
The periodogram of the \textit{TESS} data is shown in the right panel of Figure~\ref{fig:TESS_lc} up to a frequency of 0.27 mHz (equivalent to a minimum period of 60 min, which is twice the sampling rate for the \textit{TESS} data). The periodogram shows strong power at the orbital frequency and its first 8 harmonics. There are a further two potential signals located close to $3\times10^{-4}$ Hz which are not clearly associated with the orbital period. In order to investigate whether these peaks were real, noise, or related to a beat between the orbital period and the cadence of the data, a function describing the orbital modulation of the light curve was created by fitting the binned \textit{TESS} light curve shown in Figure~\ref{fig:phased_lc} using the \texttt{numpy} interpolate feature. This model light curve was then subtracted from the data, and a power spectrum of the residuals taken. This power spectrum showed no strong features, suggesting these two peaks seen on the far right of Figure~\ref{fig:TESS_lc} are not intrinsic to the system.

\subsection{\textit{XMM-Newton} OM}
\begin{figure}
    \centering
    \includegraphics[width=0.99\columnwidth]{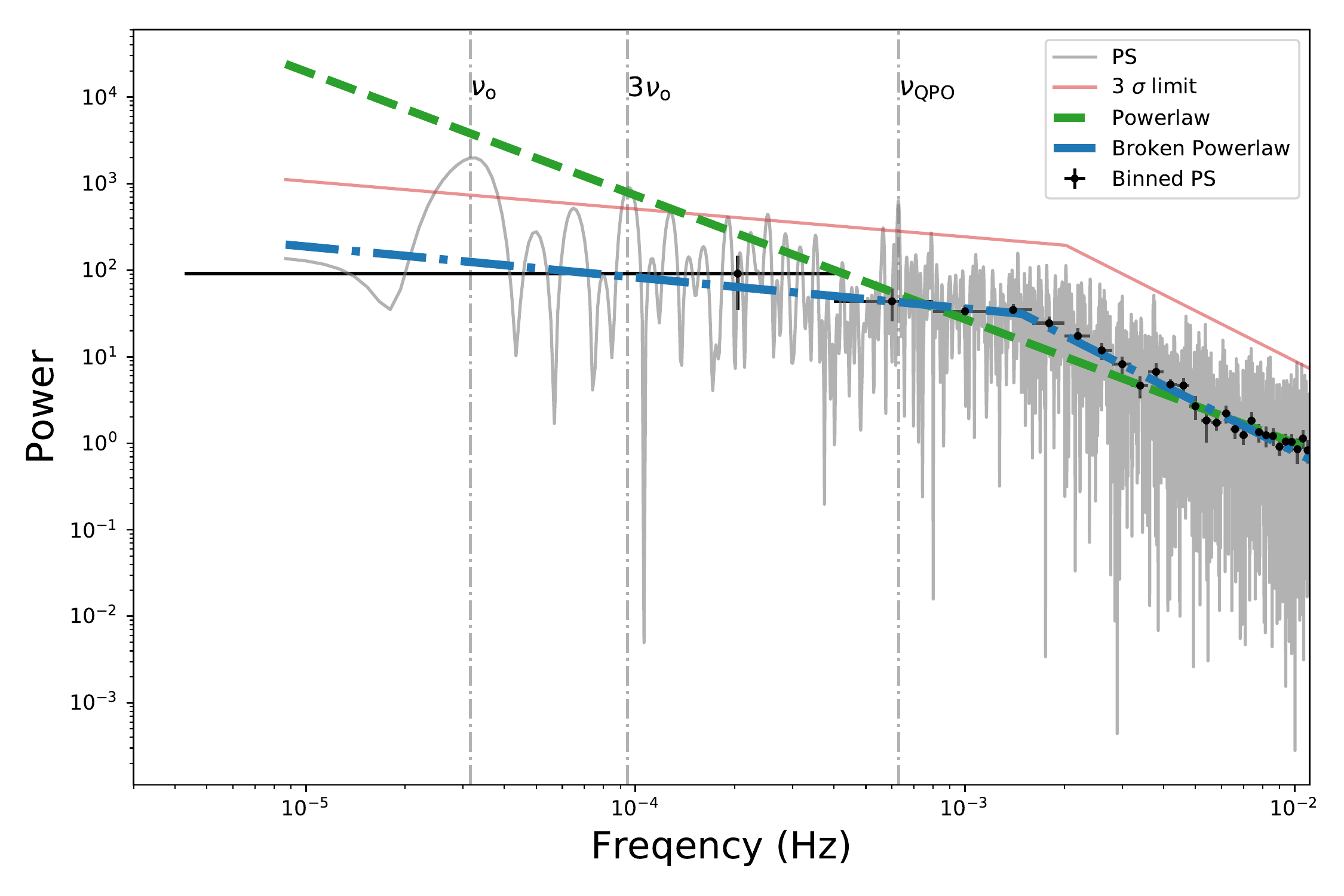}
    \caption{Power spectrum of the data obtained using the OM onboard \textit{XMM-Newton} (grey). The power spectrum was binned (black points) and fit with a power law (green, dashed line) and broken power law (blue, dashed-dotted line) model. The power spectrum was best fit by the broken power law model, with the break occurring at $1.6\times10^{-3}$ Hz. The red line represents the 3$\sigma$ level for a detected signal based on simulated light curves. There are three peaks located above this confidence level - one at the orbital frequency (marked as $\nu_{\rm{o}}$), one at three times the orbital frequency ($3\nu_{\rm{o}}$), and one at 0.63 mHz ($\nu_{\rm{QPO}}$), which corresponds to a period of 26.5 min.}
    \label{fig:OM_PS}
\end{figure}

We next took the optical light curve obtained using the OM on-board \textit{XMM-Newton} and constructed a power spectrum up to a frequency of 11 mHz (a minimum period of 1.5 min). The power spectrum is shown in Figure~\ref{fig:OM_PS}.

There are three frequencies where the detected power surpasses the $3\sigma$ level - $\nu_{\rm{O}}$ (the orbital frequency), $3\times\nu_{\rm{O}}$, and 0.63 mHz. The detection of the first two frequencies is unsurprising given that the light curve shows eclipses, but the detection of power at 0.63 mHz (equivalent to a period of 26.5 min) is surprising. One possibility is that this is related to the window function of the observations - while the length of a single observation taken with the OM lasts for 20 minutes, there is also 6 minutes of dead time between sequential observations, meaning  that the OM data has a sampling rate of 26 minutes. This could be giving rise to the signal at this period. Considering, however, that the 26.5 min period does not show up in our simulated light curves which have the same window function as the actual data, the period is likely not related to this, and is instead intrinsic to the system.

\subsection{ULTRACAM}
After removing the orbital modulation and masking the eclipse from the original light curve, the remaining data showed dramatic variability typically associated with flickering in an accretion disc. We subjected the data to a variety of time-series analysis techniques to look for short term periodicities. We treated each night of data individually to avoid the heavy aliasing which would arise had we combined all of the observations together. 

\subsection{Autocorrelation Function}\label{sec:ACF}
The autocorrelation function (ACF) for each night of data was calculated after the orbital modulation had been removed. For the data from 2017-10-16 we excluded all data taken after the eclipse so that this feature would not alter the ACF, and no gaps in the data would be present. The ACFs from each night showed correlations with a time scale of $\sim$ 2-3 min (see Figure~\ref{fig:ACF} in Appendix~\ref{sec:app_1}). The ACFs from 2017-10-14 and 2017-11-22 showed no other strong correlations. The same can not be said of the data from 2017-10-16. This ACF has not only a short time-scale correlation of about $\sim$ 2-3 min, but also a near sinusoidal correlation with a period of 21 minutes which is only coherent for the first 150 min of the observation.

\subsubsection{Power Spectrum}

\begin{figure*}
	\includegraphics[width=0.99\textwidth]{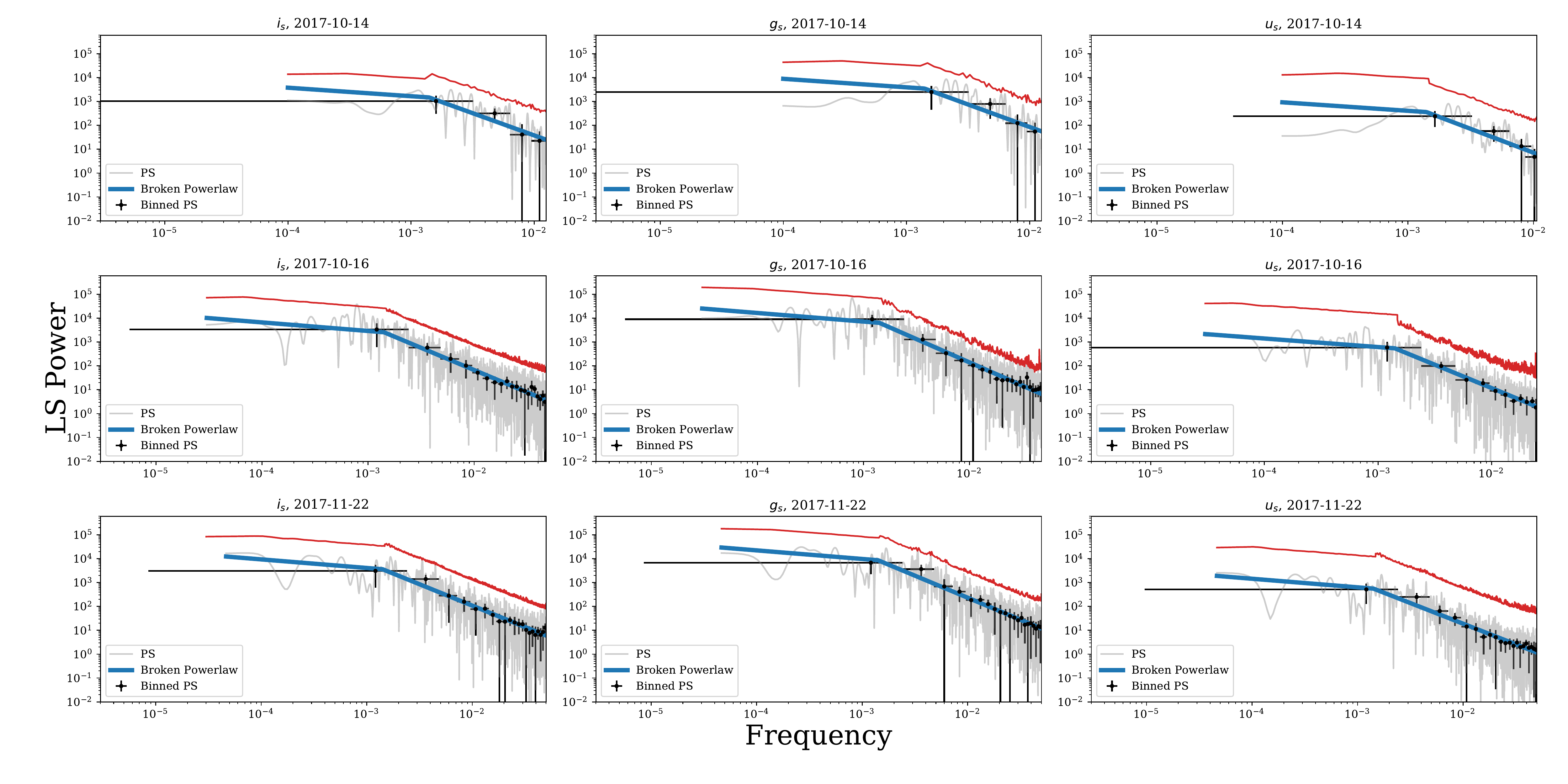}
    \caption{Lomb-Scargle periodograms for each night of data (top to bottom) in each band (left to right). The blue line is the best fit broken power law to the data, while the red line marks the 3$\sigma$ level determined as described in the text. The y-axis is dimensionless following the standard normalisation option of the Lomb-Scargle periodogram as implemented in Astropy \citep{2013A&A...558A..33A}.}
    \label{fig:full_powerspec}
\end{figure*}

Lomb-Scargle periodograms of each night of data showed strong power at varying periods between 10-30 minutes, as shown in Figure~\ref{fig:full_powerspec}. However, since the system contains an accretion disc (based on the double peaked optical emission lines reported in \citetalias{2016ApJ...831...89S}), we expect the power spectra to be dominated by pink noise. This is correlated noise with a power-law spectrum $f^{\alpha}$, with $\alpha=-1$, and is not unusual in systems which are dominated by ``flickering'' \citep{1987Natur.325..694L}. The presence of pink noise makes the determination of the significance of any peaks in a power spectrum difficult, as the usual metrics such as the False Alarm Probability depend on the noise in the data being uncorrelated ($\alpha=0.0$). As such, the 3$\sigma$ level for whether a period was real or a noise peak for each band and each night was found using the method discussed in Appendix~\ref{sec:app_1}.

The only power spectrum which had a peak detected above the 3$\sigma$ level was from the data taken on 2017-10-16 in the $i_s$, and the peak was found to lie at $\sim$21 min. While there is a peak in the  $g_s$ and $u_s$ light curve around the same period, it is not detected at the 3$\sigma$ level. This may be due to a combination of the lower temporal sampling and lower S/N of the $u_s$ and the lower S/N of the $g_s$ data when compared to the $i_s$ data.

\subsection{Gaussian Process Modelling}\label{sec:GPM}
We next attempted to investigate the nature of the flickering using Gaussian Process Modelling (GPM) methods. A Gaussian Process is a collection of random variables, any finite number of which have a joint Gaussian distribution defined by a certain covariance matrix. While traditionally the covariance matrix is calculated from the data, in GPM the matrix is constructed using covariance kernels which attempt to model the correlations between data points. We tested two different kernels when analysing the optical data presented here. The first kernel was the sum of a a Mat\'{e}rn covariance kernel to allow for covariances between data points on a length scale of $\ell_1$ and a periodic kernel which allowed for variations on a time scale equal to the orbital period. The second kernel had a third component to allow for short-term periodic variations in addition to these two components of the first kernel. A detailed discussion of the kernels used in this analysis is included in Appendix~\ref{sec:app_1}.

The kernels were implemented using \textsc{scikit-learn} \citep{scikit-learn} and the hyper-parameters were tuned using a Markov Chain Monte Carlo (MCMC) sampler implemented using \textsc{emcee} \citep{2013PASP..125..306F}. As with our light curve modelling, $P_{orb}$ was fixed to 0.3667200 d. The kernels were initially trained on the combined i$_s$ data from 2017-10-14 and 2017-10-16 after converting the data from magnitude to flux. For both of the kernels, we found that length scale over which data was correlated was $2.6\pm0.3$ min, as suggested by the ACF in Section~\ref{sec:ACF}. While a solution was found using the second kernel with the short time scale periodicity, the marginal log likelihood never exceeded the marginal log likelihood of the first kernel. Even when we limited our analysis to just the data taken on 2017-10-16 prior to the eclipse (the same data used when calculating the middle row of panels of Figure~\ref{fig:full_powerspec}), we again find that the simpler first kernel is more likely than the second, suggesting that the periodicity detected using the Lomb-Scargle periodogram is transient in nature.

\begin{figure}
	\includegraphics[width=0.99\columnwidth]{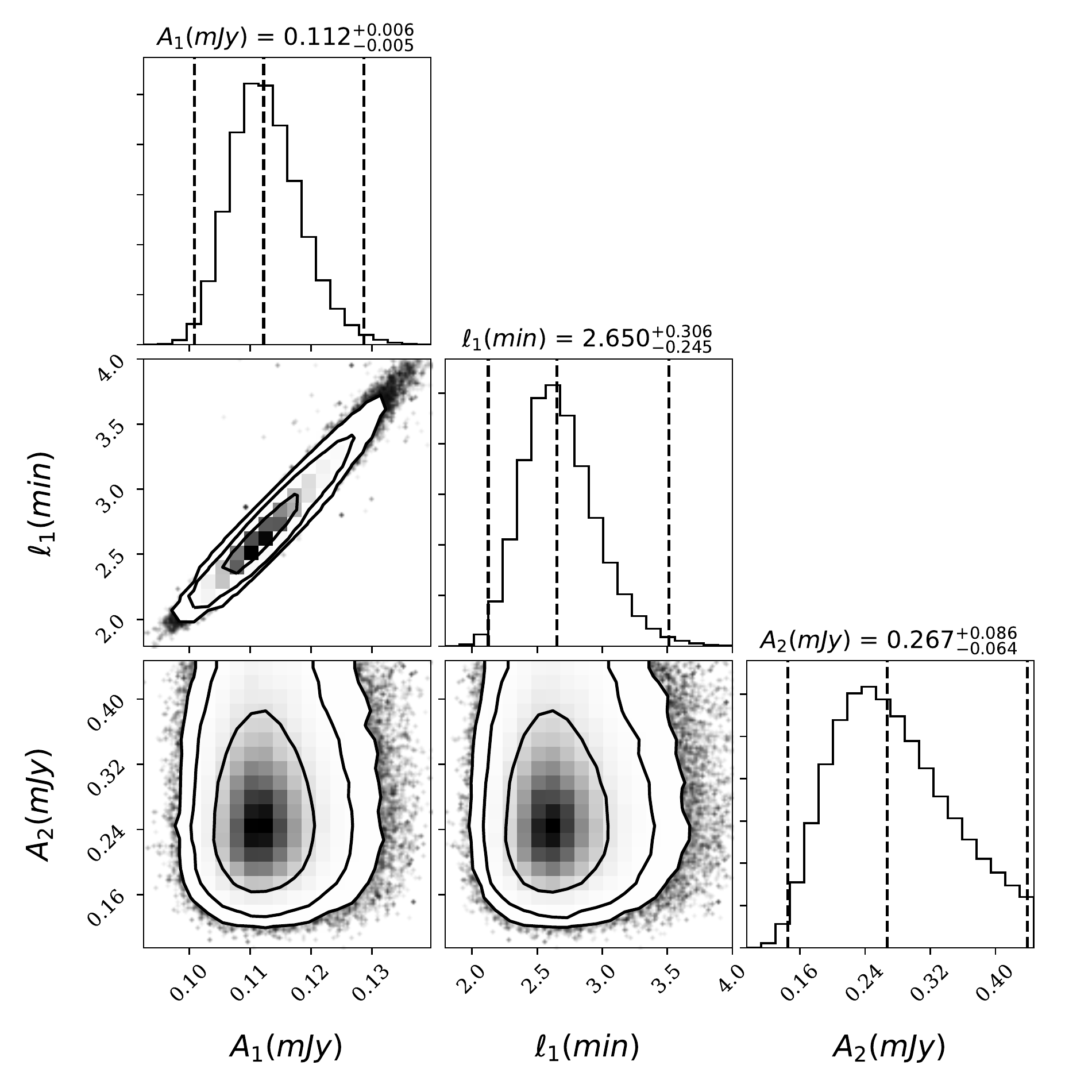}
    \caption{Results of the MCMC analysis of our data using Gaussian Process modelling. The modelling shows that the data has a short term correlation of 2.6 min ($\ell$) with an amplitude of close to 0.1 mJy ($A_1$). The orbital modulation has an amplitude of 0.26 mJy ($A_2$).}
    \label{fig:GPM_MCMC}
\end{figure}

The best fitting $A_1$, $\ell$, and $A_2$ are given in Figure~\ref{fig:GPM_MCMC}. Due to the very short time-scale correlation in our kernel of 2.7 minutes, the model quickly loses the ability to accurately predict fluxes outside of the observed data, limiting the use of this technique.

\section{X-ray Spectroscopy}
\begin{figure}
	\includegraphics[width=0.99\columnwidth]{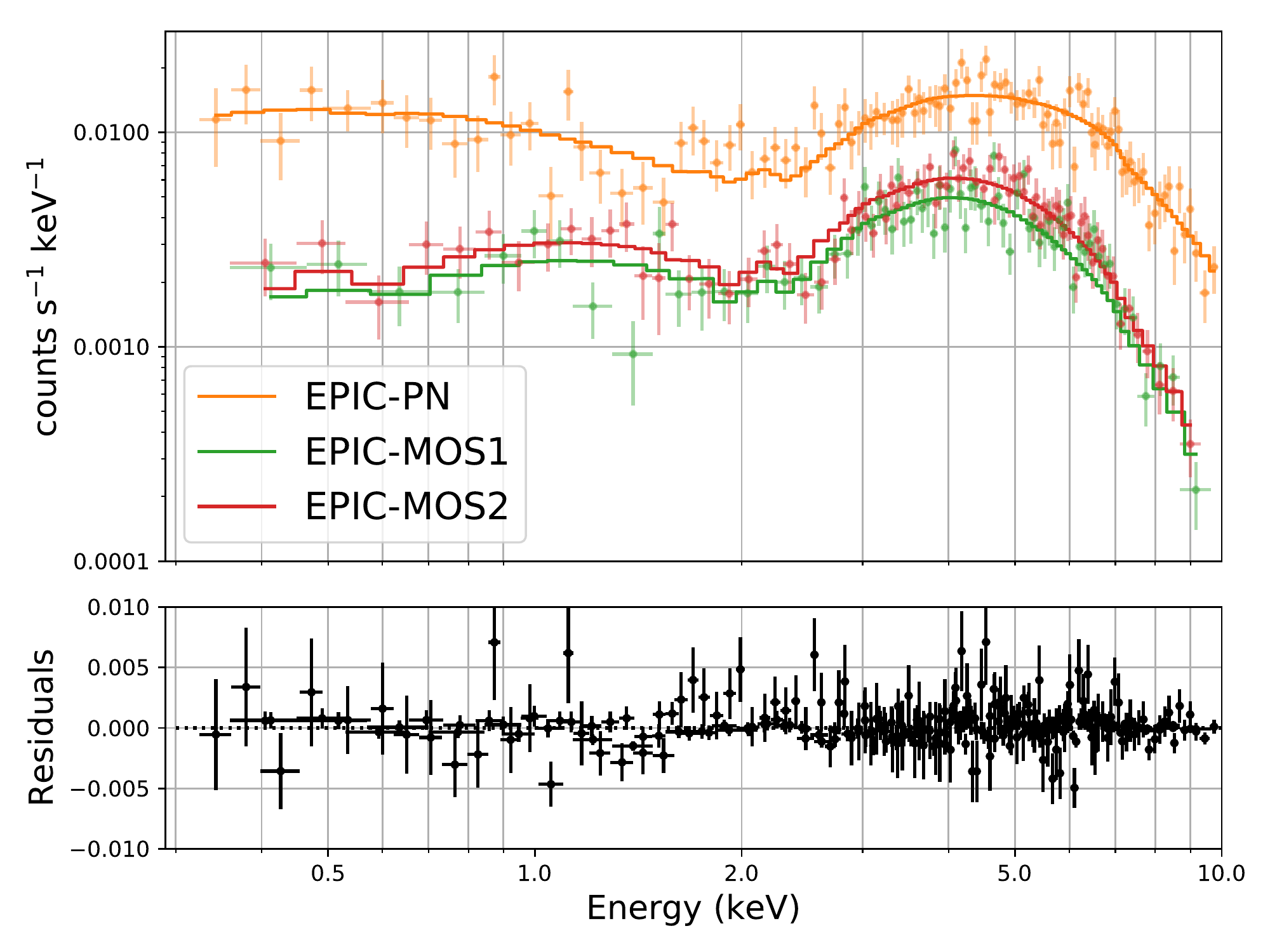}
    \caption{The X-ray spectra from 0.3 - 10 keV obtained using the EPIC-PN, -MOS1, and -MOS2 instruments onboard \textit{XMM-Newton}, alongside the best-fit partially absorbed power law model.}
    \label{fig:Xray_spec}
\end{figure}

The extracted X-ray spectra covering 0.3-10 keV from the PN and both MOS1 and MOS2 instruments onboard \textit{XMM-Newton} are shown in Figure~\ref{fig:Xray_spec}. We modelled the X-ray emission using \texttt{Xspec} v. 12.10.1 \citep{Xspec}. We initally began with a simple absorbed power law (\textsc{const$\:\times\:$tbabs$\:\times\:$powerlaw}, where the \textsc{const} component was included to allow for differences between the PN and MOS instruments and \textsc{tbabs} is the Tuebingen-Boulder ISM absorption model; \citealt{2000ApJ...542..914W}). While this model was able to describe the hard X-ray tail of the spectrum (energies $>2$ keV) it could not fit the spectrum at soft (0.1-2 keV) energies, with a $\chi^2$ value of 756.656 for 260 degrees of freedom. Since this source is edge on, we expect the amount of absorption due to material within the binary system to be high. As such, we next fit a partially absorbed powerlaw to the data to account for the circumstellar absorption. The model (\textsc{const$\:\times\:$tbabs$\:\times\:$pcfabs$\:\times\:$powerlaw}) fit the data well, with a $\chi^2=272.928$ for 259 degrees of freedom. The model is shown alongside the spectra in Figure~\ref{fig:Xray_spec}, and the best fit parameters are given in Table~\ref{tab:xray_model}.

We also tried fitting the spectrum with an additional Gaussian emission component to account for any emission at the 6.4 keV \ion{Fe}{K$\alpha$} emission line. The addition of such a component did not increase the goodness-of-fit by a statistically significant amount.

The best fit parameters show that the interstellar absorber accounts for very little absorption in the spectrum, with only an upper limit on the density calculable. Nearly all of the absorption arises due to the high density and covering fraction of the partial absorber, indicative of significant absorption by the accretion disc. This is in line with the high inclination derived from the optical light curve modelling.

We computed the unabsorbed 2-10 keV flux by fitting the spectrum with the \texttt{Xspec} model \textsc{cflux} and using the best-fit parameters found for the above model. The unabsorbed 2-10 keV flux was found to be $(1.73\pm0.08)\times10^{-12}$ erg cm$^{-2}$ s$^{-1}$.

\begin{table}
	\centering
	\caption{Best fit parameters from an absorbed powerlaw fit to the X-ray Spectrum}
	\begin{tabular}{r r c}
		\hline
		Component	& Parameter		& Value\\
		\hline\hline
        tabs        & $n_{\rm{H}}$ (cm$^{-2}$)  & <1$\times10^{19}$\\
        pcfabs      & $n_{\rm{H}}$ (cm$^{-2}$)  & $(9\pm1)\times10^{22}$\\
                    & CovFrav                   & $0.96\pm0.01$\\
        powerlaw    & $\Gamma$                  & $1.3\pm0.1$\\
                    & norm$^{a}$                & $(1.9\pm0.8)\times10^{-4}$\\
		\hline
        \multicolumn{3}{l}{$^{a}$photons keV$^{-2}$ cm$^{-2}$ s$^{-1}$ at 1 keV}\\
	\end{tabular}
	\label{tab:xray_model}
\end{table}

\section{Discussion}

\subsection{Component Masses}
\begin{figure}
	\includegraphics[width=0.99\columnwidth]{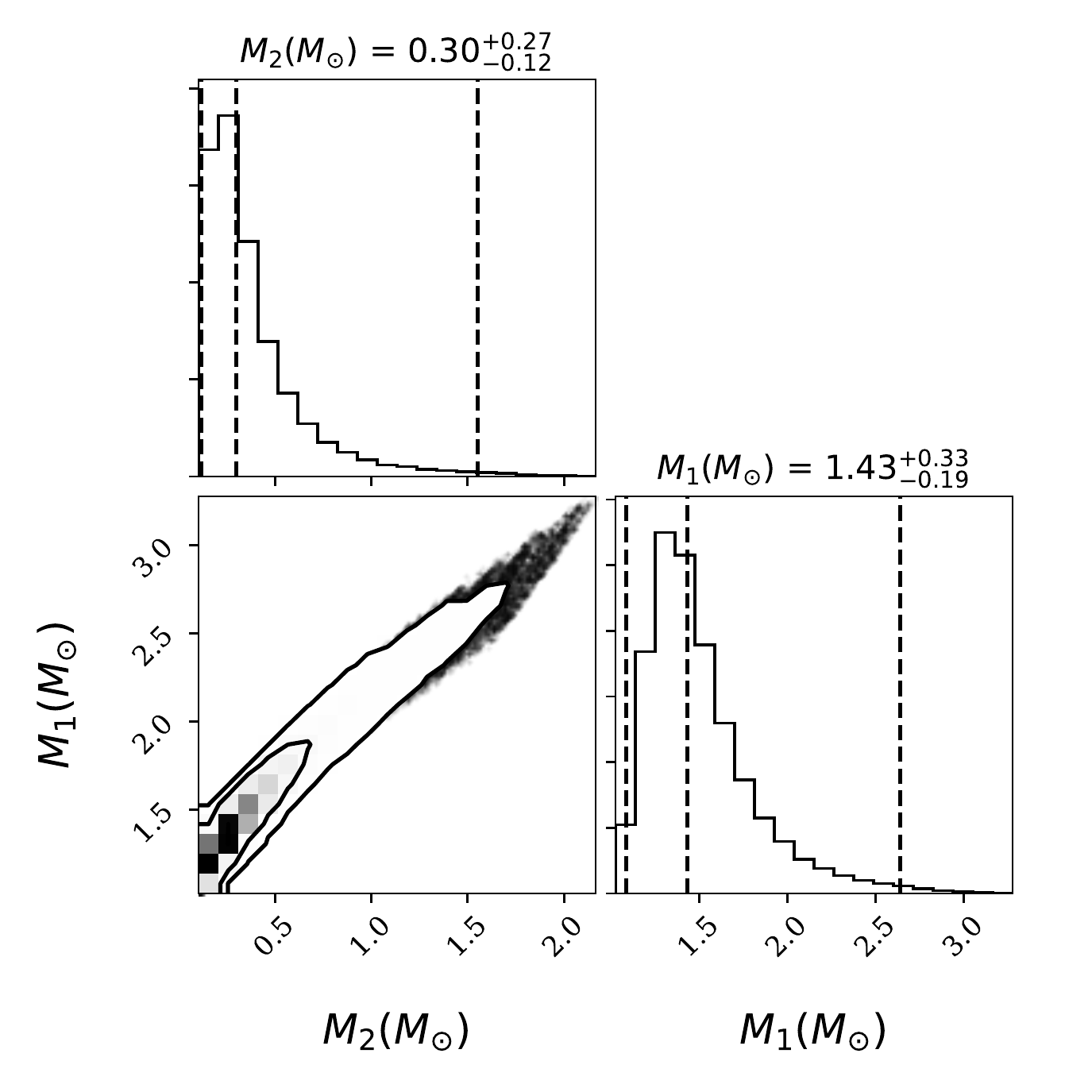}
    \caption{Component masses of system. The upper limit on the primary mass is close to the theoretical limit on the mass of a neutron star. Contours are at the 2$\sigma$ and 3$\sigma$ levels.}
    \label{fig:masses}
\end{figure}

For each model generated by ELC, the software also computes the individual masses using the system inclination, mass ratio, orbital period, and $K_{2}$. The primary and secondary masses for every model computed in our analysis are shown in Figure~\ref{fig:masses}. The masses are not very well constrained, with $M_1=1.43^{+0.33}_{-0.19}$ M$_{\odot}$ and $M_2=0.30^{+0.27}_{-0.12}$ M$_{\odot}$ at the 1$\sigma$ level. We can use the lower bound of $K_2 > 283$ km s$^{-1}$ and the measured $P_{\rm{orb}}$ to put a lower bound on the primary mass of $M_1>0.86$ M$_{\odot}$ by using the mass function formula
\begin{equation}
M_1 = \left(\frac{1}{q}+1\right)^2 \frac{P_{\rm{orb}} \: K_2^3}{2 \pi \: G \sin{i}^3},
\end{equation}
and letting $q \to \infty$ (for $q=\frac{M_1}{M_2}$ as defined for ELC) and $i \to 90$\degree. From our modelling, the primary mass is less than $2.5$ M$_{\odot}$ at the 3$\sigma$ level. However, this upper bound is not entirely reliable, as the observed $K_2$ may be biased by irradiation of the secondary. Figure~\ref{fig:masses} also shows why there is a large error on the mass ratio from our modelling - for a very low-mass secondary ($<0.18$ M$_{\odot}$), the mass ratio of the system can be as large as 10. If we require that the secondary mass $>0.18$ M$_{\odot}$, then the mass ratio is more tightly constrained to $q=3.5\pm1.0$.

We note that our constraints on the individual masses are not as tight as those derived by \citetalias{2016ApJ...831...89S}. This is because \citetalias{2016ApJ...831...89S} measured the radial velocity feature of the emission lines and assumed that this velocity was equivalent to the velocity of the primary star. This assumes that the accretion disc extends down to the surface of the primary, which is not entirely obvious. Additionally, any brightness asymmetries in the accretion disc would cause an incorrect measurement of $K_1$. Due to these two points, we did not include this value in our modelling, meaning the mass ratio was left unconstrained. A direct measurement of $K_{1}$ would become possible if the primary were detected as a radio or X-ray pulsar.

\subsection{Is \fgl\ a cataclysmic variable?}
The mass of the primary in \fgl\ has still not been well constrained, and the system could still be a cataclysmic variable (an interacting binary system with a white dwarf primary) with a particularly heavy white dwarf primary. Below we lay out the arguments that suggest the primary is not a white dwarf.

First, the optical light curve of the system helps rule out a typical dwarf nova classification. This type of cataclysmic variable exhibits outbursts related to instabilities in their accretion discs which have a typical amplitude of several magnitudes at optical wavelengths. The lack of any observed outbursts in \fgl\ in our data or in the data presented in \citetalias{2016ApJ...831...89S} help immediately rule out this classification.

The observed parallax of \fgl\ in the recent Gaia 2nd data release is $0.38\pm0.07$ mas \citep{2018A&A...616A...1G}. This implies a distance to \fgl\ of $1.8<d<5.6$ kpc at the $3\sigma$ level, estimated using the methods described in \citet{2018arXiv180409376L} and with a length scale of 1 kpc for the prior. The distance to \fgl\ combined with the 0.3-10 keV X-ray flux of $2.0^{+15.9}_{-0.5}\times 10^{-12}$ erg cm$^{-2}$ s$^{-1}$ measured here suggests that the 0.3-10 keV X-ray luminosity of the source is $7\times 10^{32}<L_{\rm{X}}<8\times 10^{33}$ erg s$^{-1}$. Such a high X-ray luminosity is incompatible with dwarf novae cataclysmic variables, which typically have X-ray luminosities $~10^{29-30}$ erg s$^{-1}$. However, cataclysmic variables which have persistently high mass accretion rates (called nova-likes) and cataclysmic variables with magnetic white dwarfs have much higher X-ray and optical luminosities than their dwarf nova cousins. In particular, intermediate polars (which are magnetic CVs that harbour accretion discs; IPs) have X-ray luminosities of $10^{32-33}$ erg s$^{-1}$ \citep{2016ApJ...818..136X}, which is consistent with the lower limit on the X-ray luminosity of \fgl. Additionally, an IP at same distance as \fgl\ would have a similar apparent magnitude at optical wavelengths as \fgl. However, we note that the above comparisons are lower limits, since we have assumed the minimum distance possible to \fgl, and it is likely that the X-ray flux from \fgl\ is much higher than the lower limit implies (as hinted at by the ELC modelling). The optical polarimetry does not help rule out a cataclysmic variable classification, as the upper limit of 1\%\ on circular polarisation is higher than the detected circular polarisation in most magnetic CVs \citep{2009A&A...496..891B}.

The most convincing evidence for ruling out a white dwarf primary is the association of \fgl\ with a GeV $\gamma$-ray source, as no cataclysmic variable has ever been observed to produce such high energy photons in their regular states. The only state in which a system with a white dwarf primary has exhibited $\gamma$-ray emission is when a CV undergoes a nova eruption due to thermonuclear burning on its surface (for example V407 Cygni; \citealt{2010Sci...329..817A}), and the $\gamma$-rays are only produced during the eruption itself. The confirmation of the persistent $\gamma$-ray eclipse in \fgl\ proves that the $\gamma$-rays are coming from close to or directly from the primary in this system, and are not due to a nova eruption.

\subsection{Is \fgl\ a transitional MSP?}
If \fgl\ is not a CV, then it is a low-mass X-ray binary (LMXB) containing either a NS or a black hole primary. The detection of radio or $\gamma$-ray pulsations from the primary or the detection of an X-ray burst would confirm the existence of a NS primary, while an accurate mass measurement could be used to distinguish between these two primaries. Regarding the first point, \fgl\ has yet to be detected as a persistent or periodic radio source, and regarding the X-ray flaring, the X-ray light curve shows significant variability (even out side times of high background, as shown in Figure~\ref{fig:xray_lc}). In particular, there is a single event where the count rate increases by a factor of 5. However, these events are not resolved with the 100s bin width which is required to obtain an adequate S/N in the light curve, making the actual duration and shape of the events impossible to determine.

As such, we must rely on the mass constraints given in this paper to determine the nature of the compact object. The maximum mass of the primary star is 2.42 M$_{\odot}$ at the 3$\sigma$ level, which is close to the theoretical upper limit for the mass of a NS. This means the most likely scenario is that the primary is not a black hole, but a NS. Given this, the question becomes is the system a regular LMXB with a NS primary, or a tMSP in an accreting state?

tMSPs have a combination of optical features and X-ray features which can be used to distinguish them from regular LMXBs - optical/X-ray flares and a bi-modality in the optical/X-ray flux distribution (see \citealt{2015ApJ...806..148B} for an example of X-ray bi-modality and flares, \citealt{2015MNRAS.453.3461S} for an example of optical bi-modality, and \citealt{2018MNRAS.477.1120K} for examples of optical flares). The origin of these features is still uncertain, and while there are a multitude of models which attempt to explain the features (\citealt{2014ApJ...795...72L}; \citealt{2015ApJ...807...33P}; \citealt{2019arXiv190410433P}), most of them agree that the features likely arise from interactions deep within the accretion disc, close to the magnetosphere of the NS.

\fgl\ shows no such bi-modality in its optical flux distributions (after removal of the orbital modulation, the flux distributions in all 3 bands were log-normally distributed), or in the X-ray light curve obtained with XMM-Newton (with the caveat that a majority of the X-ray light curve must be ignored due to high background). Such a discrepancy is not at odds with the classification of \fgl\ as a tMSP, as \citet{2018MNRAS.477.1120K} showed that in long term optical observations of PSR J1023+0038 taken in its accretion states there were long periods of time when the flux of PSR J1023+0038 did not show a bi-modal distribution. 

It may also be that our ability to detect bi-modality is inclination dependent. Since the inner part of the disc is where the bi-modality supposedly originates, and \fgl\ is an edge-on system, the region where this bi-modality arises may be hidden from us. This allows for the classification of \fgl\ as a tMSP despite the non-detection of a bi-modal optical and X-ray flux distribution.

Additionally, the derived X-ray flux from the \textit{XMM-Newton} observations and the measured power law index of $\Gamma=1.3\pm0.1$ are consistent with a tMSP in an accreting state. The distance limits on \fgl\ combined with the measured 2-10 keV flux provide limits on the X-ray luminosity of \fgl\ as $7\times 10^{32}<L_{\rm{X}}<8\times 10^{33}$ erg s$^{-1}$. The measured X-ray luminosity of PSR J1023+0038 during its accreting state is $3\times10^{33}$ erg s$^{-1}$ (when in the X-ray high mode). If \fgl\ were to have a similar luminosity, then it would be located at 4 kpc. The next \textit{Gaia} data release should have a more precise parallax for \fgl\, which will then better constrain the X-ray luminosity.

Finally, typical accreting LMXBs do not have $\gamma$-ray counterparts, while the 3 confirmed tMSPs (PSR J1023+0038, IGR J18245-2452, and PSR J1227-4853) and 1 of the potential tMSPs (3FGL J1544.6-1125) do. While the definite detection of a $\gamma$-ray eclipse in \fgl\ shown in this paper strongly supports the claim by \citetalias{2016ApJ...831...89S} that \fgl\ is a tMSP in an accreting state, the conclusive confirmation of J0427 as an accretion-state tMSP requires the
detection of millisecond pulsations from the spinning neutron star primary.

\subsubsection*{Pulsation detection feasibility}
While radio pulsations have only been detected from tMSPs in rotation-powered states, optical and X-ray pulsations have been detected from PSR~J1023$+$0038 in its accreting state (\citealt{2015ApJ...807...62A}; \citealt{2016ApJ...830..122J}; \citealt{2017NatAs...1..854A}; \citealt{2019MNRAS.485L.109Z}). Additionally, while $\gamma$-ray pulsations have been detected from a large number of nearby MSPs, it remains an open question as to whether or not $\gamma$-ray pulsations from a tMSP are also suppressed during accreting states. Of the three confirmed tMSPs, $\gamma$-ray pulsations have only been detected from PSR~J1227$-$4853 during its rotation powered state \citep{2015ApJ...806...91J}, but the low signal-to-noise ratio and unpredictable variations in the orbital period have prevented the extrapolation of the timing solution back to the pre-transition epoch to check for pulsations in the accreting state.

For \fgl\, whose pulsation period remains unknown, the detection of $\gamma$-ray pulsations would require a multidimensional blind search over several timing parameters. Given sufficiently precise knowledge of the orbital parameters (period, phase and projected semi-major axis), such searches are capable of discovering binary MSPs \citep{2012Sci...338.1314P}. However, for \fgl\ the current parameter constraints (from the orbital ephemeris of Equation~\ref{eq:ephemeris}, and the semi-major axis range of $0.7\,\textrm{s} \lesssim (a\sin i)/c \lesssim 1.6 \textrm{s}$ inferred from the parameters in Table~\ref{tab:model_lc_results}) still leave a prohibitively large volume to search, and even using e.g. the thousands of computers participating in the \textit{Einstein@Home} volunteer computing project (\citealt{2013ApJ...773...91A}; \citealt{2017ApJ...834..106C}), a full blind search would take more than a year to complete (L. Nieder, private communication). Given that it is still unclear whether or not a tMSP in an accreting state will even emit detectable $\gamma$-ray pulsations, we do not believe that such a search would currently be a good use of computing resources, but we note that this may change in the future should the orbital ephemeris be further refined with additional optical observations.

\section{Conclusions}

We have presented further optical, X-ray, and $\gamma$-ray observations of \fgl. The $\gamma$-ray eclipse is now detected with a significance $>5\sigma$. This confirms that the $\gamma$-ray emission is associated with the X-ray and optical source, establishing \fgl's membership as one of a rare class of accreting binaries. The high time resolution optical data show rapid flickering on a timescale of 2.4 mins, with hints of an underlying 21 min period. Modelling of the optical light curve has placed tight constraints on the inclination of the system to be $84\pm3$\degree, and we find that there is significant evidence for heating of the secondary star by the primary. 

We do not find evidence for bi-modality in the optical or X-ray light curvs. This is still consistent with \fgl\ belonging to the tMSP class of objects, as tMSPs do not always show bi-modality in their optical and X-ray light curves, and our ability to detect bi-modality may be inclination-dependant, with the bi-modality of high inclination systems such as \fgl\ difficult to detect due to obscuration of the region associated with this behaviour by the outer parts of the accretion disc. While we have not been able to rule out a white dwarf primary by modelling the optical light curve, it is likely that the primary is too heavy to be a white dwarf, and the now significant detection of a $\gamma$-ray eclipse makes any primary other than a NS in a tMSP difficult to explain.

Definitive classification of \fgl\ as a tMSP requires detection of radio/optical/X-ray pulsations from the primary, or the detection of a state transition. One thing that is beyond debate is that additional optical, X-ray, radio, and $\gamma$-ray data should be obtained. In addition to searching for state transitions, further optical photometry can be used to confirm the $\sim21$ min period, and optical spectroscopy would allow the use of Doppler tomography and spectral eclipse mapping to better understand the accretion structures within this system.

\section*{Acknowledgements}

The authors thank Guillaume Voisin, Anna Scaife, Matthew Green for useful discussions regarding Gaussian Process Modelling. M.R.K is funded through a Newton International Fellowship provided by the Royal Society. R.P.B., C.J.C., D.M-S., and J.S. acknowledge support from the ERC under the European Union's Horizon 2020 research and innovation programme (grant agreement No. 715051; Spiders). VSD and ULTRACAM are supported by the UK's Science and Technology Facilities Council (STFC). The research by DAHB and SBP is supported by the National Research Foundation of South Africa. Based on observations collected at the European Organisation for Astronomical Research in the Southern Hemisphere using the New Technology Telescope under ESO programmes 0100.D-0425(B) and 0102.D-0953(A). This research made use of Astropy, a community-developed core Python package for Astronomy (\citealt{2013A&A...558A..33A}; \citealt{2018AJ....156..123A}). This research also made use of celerite, a package for scalable Gaussian Process regression \citep{celerite}. This paper includes data collected by the \textit{TESS} mission. Funding for the TESS mission is provided by the NASA Explorer Program. This work is based on observations obtained with \textit{XMM-Newton}, an ESA science mission with instruments and contributions directly funded by ESA Member States and NASA. Work at NRL is supported by NASA.

The \textit{Fermi} LAT Collaboration acknowledges generous ongoing support
from a number of agencies and institutes that have supported both the
development and the operation of the LAT as well as scientific data analysis.
These include the National Aeronautics and Space Administration and the
Department of Energy in the United States, the Commissariat \`a l'Energie Atomique
and the Centre National de la Recherche Scientifique / Institut National de Physique
Nucl\'eaire et de Physique des Particules in France, the Agenzia Spaziale Italiana
and the Istituto Nazionale di Fisica Nucleare in Italy, the Ministry of Education,
Culture, Sports, Science and Technology (MEXT), High Energy Accelerator Research
Organization (KEK) and Japan Aerospace Exploration Agency (JAXA) in Japan, and
the K.~A.~Wallenberg Foundation, the Swedish Research Council and the
Swedish National Space Board in Sweden.
 
Additional support for science analysis during the operations phase is gratefully
acknowledged from the Istituto Nazionale di Astrofisica in Italy and the Centre
National d'\'Etudes Spatiales in France. This work performed in part under DOE
Contract DE-AC02-76SF00515.




\bibliographystyle{mnras}
\bibliography{j0427}



\appendix

\section{Time series analysis}\label{sec:app_1}
\subsection{Estimating the 3$\sigma$ values for signal detection in a periodogram}
In order to construct accurate confidence levels for peaks in the power spectrum, it is important to know what the underlying shape of the power spectrum is. Typically power spectra are modelled as power laws ($P(f)\propto f^{\alpha}$ where $\alpha$ is the spectral index). A flat spectral index ($\alpha \sim 0$) suggests that the light curve is dominated by white noise, while a steep spectral index ($\alpha \sim -2$) suggests strong correlated red noise is present in the data. The light curves of systems which contain accretion discs often are often dominated by ``flickering'', which shows up in power spectra as pink noise (a power-law spectrum with $\alpha=-1$; \citealt{1987Natur.325..694L}). For each of the power spectra presented in this paper, we first binned the power spectrum in question such that the distribution of power within each bin followed a log-normal distribution, and then fit both a power law and a broken power law to the binned data.

Once the frequency dependence of the power spectrum was measured, the 3$\sigma$ level for identifying significant periods present in the power spectrum at each frequency could be estimated. This was done by generating 100,000 light curves which had the same temporal sampling as our data, but were generated using a fake power spectrum with either the best fitting power law or a broken power law from the first step, and then creating the light curve using the algorithm described in \citet{1995A&A...300..707T} and implemented in \textsc{Stingray}\footnote{\textsc{Stingray} is a \textsc{Python} package for X-ray astronomy, and is available at \url{https://github.com/StingraySoftware/stingray}}. The power spectrum of each of these fake light curves was taken, and the power at each frequency recorded. The distribution of powers at each frequency were then fit with a cumulative distribution function assuming the noise is Gaussian distributed (equation 53 of \citealt{2018ApJS..236...16V}), and the 3$\sigma$ level at each frequency was taken to be the central value of the Gaussian plus 3 standard deviations. The threshold line is not the ``single trial'' threshold, but shows the 3-sigma level after accounting for the number of independent frequencies in the power spectrum. In this case, this number was assumed to be the length of the frequency range of the power spectrum up to the Nyquist frequency ($f_{\rm Ny} = 0.5 \: t_{\rm samp}$ where $t_{\rm samp}$ is the sampling rate) divided by a delta frequency such that the power spectrum was Nyquist sampled ($\delta f = 1/T_{\rm obs}$ where $T_{\rm obs}$ is the length of the observations). The results of this error estimation are plotted as the 3$\sigma$ levels in each of the power spectra in this paper, and code to reproduce these plots is hosted online\footnote{\url{https://github.com/mkenne15/papers/tree/master/4FGLJ0427}}.

\begin{figure*}
	\includegraphics[width=0.99\textwidth]{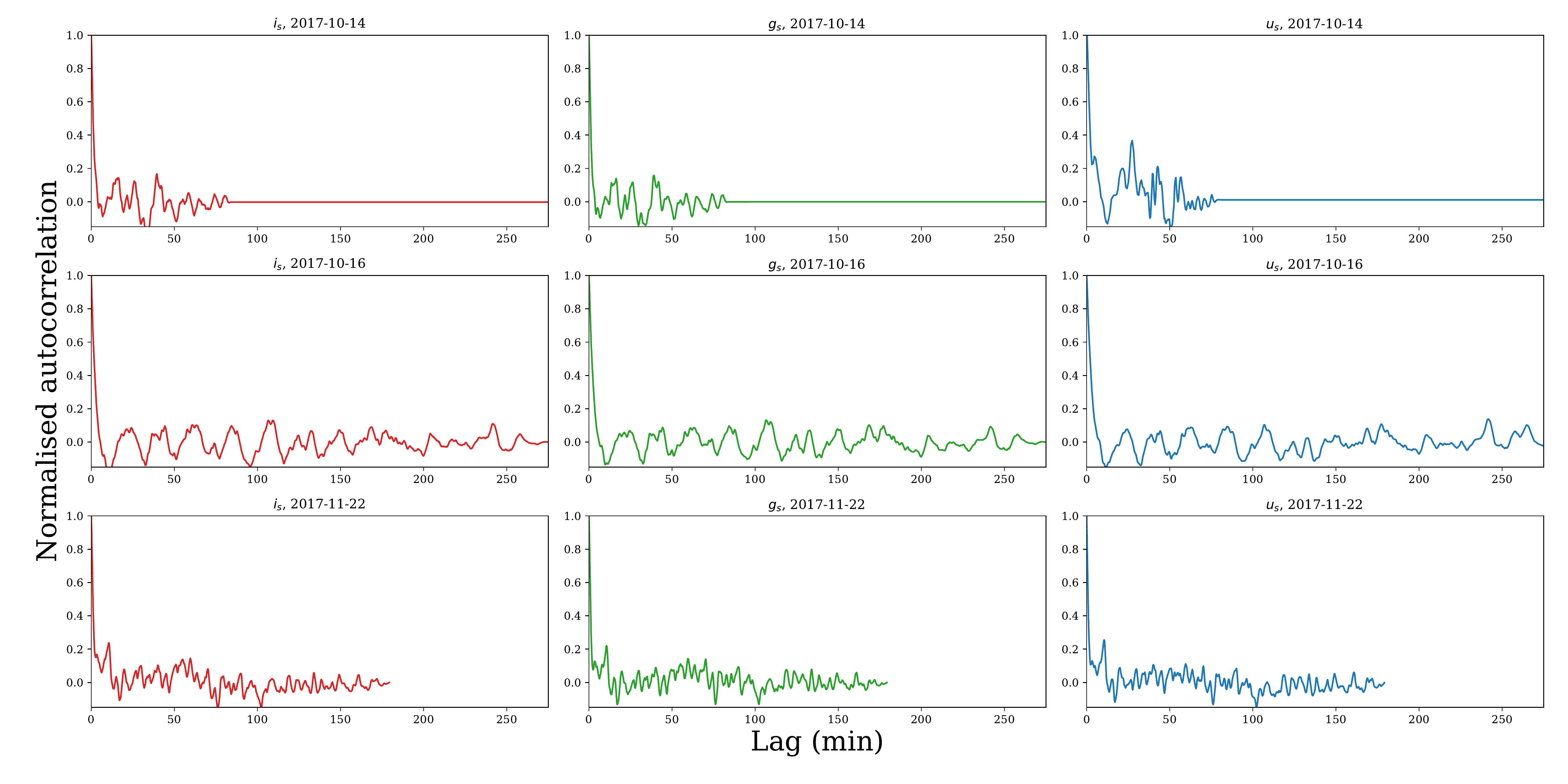}
    \caption{The auto correlation function for each night of data (top to bottom) in each band (left to right). Each night of data shows correlations on a time scale of 2-3 min, while the data from 2017-10-16 also show weak periodic correlations with a time scale of $\sim20$ min.}
    \label{fig:ACF}
\end{figure*}

\subsection{Gaussian Process Modelling of the optical light curve}
For a pair of data points, \((x_i,x_j)\), we tested two kernels: one of the form

\begin{equation}\label{e:GPM1}
\begin{split}
k(x_i, x_j) = {} & (A_1)^2\mathrm{exp}\left(-\frac{(x_i-x_j)}{\ell_1}\right)+\\
				 & (A_2)^2\mathrm{exp}\left(-\frac{2}{(\ell_2) ^2}\sin ^2 \left( \frac{\pi (x_i-x_j)}{P_{orb}} \right)\right),
\end{split}
\end{equation}

and another of the form

\begin{equation}\label{e:GPM2}
\begin{split}
k(x_i, x_j) = {} & (A_1)^2\mathrm{exp}\left(-\frac{(x_i-x_j)}{\ell_1}\right) +\\
          		 & (A_2)^2\mathrm{exp}\left(-\frac{2}{(\ell_2) ^2}\sin ^2 \left( \frac{\pi (x_i-x_j)}{P_{orb}} \right)\right)+\\
                 & (A_3)^2\mathrm{exp}\left(-\frac{2}{(\ell_3) ^2}\sin ^2 \left( \frac{\pi (x_i-x_j)}{P} \right)\right).
\end{split}
\end{equation}

The first term in each kernel is a Mat\'{e}rn covariance function with $\nu=0.5$ and allows for covariances between data points on a length scale of $\ell_1$ \citep{2006gpml.book.....R}, the second term in each kernel allows for periodic variations on a time scale equal to the orbital period, and the third term in the second kernel was to allow for short-term periodic variations. $\ell_2$ is defined such that $\ell_2<1$ allows for rough periodic variations, while $\ell_2>1$ lets the variations be more strictly sinusoidal in their appearance (the same is true of $\ell_3$). We fixed $\ell_2$ and $\ell_3$ to 1.0 based on the autocorrelation function calculated previously in Section~\ref{sec:ACF}. In GPM, $A_1$, $A_2$, $\ell_1$, $\ell_2$, and $P$ are often referred to as hyper-parameters, as their physical meanings can, at times, be difficult to understand.



\bsp	
\label{lastpage}
\end{document}